\documentclass{aa}  
\usepackage[varg]{txfonts}
\usepackage{graphicx}
\usepackage{natbib}

\newcommand{\fig}[1]{Fig.~\ref{#1}}
\newcommand{\eq}[1]{Eq.~\ref{#1}}

\begin{document}

\title{Prompt thermal emission in gamma-ray bursts}

\date{Accepted for publication in A\&A, January 25, 2013}
\author{
R. Hasco{\"e}t\thanks{email: hascoet@iap.fr}\fnmsep\thanks{Currently at: Physics Department and Astronomy Department, Columbia University, 538 West 120th Street, New York, NY 10027, USA}
          \and
          F. Daigne\thanks{Institut Universitaire de France}
          \and
          R. Mochkovitch
          }
\authorrunning{R. Hasco\"et~et~al.}
\titlerunning{Prompt thermal emission in Gamma-Ray Bursts}

\institute{UPMC-CNRS, UMR7095, Institut d'Astrophysique de Paris, F-75014, Paris, France}

\abstract
{
GRB spectra 
appear non-thermal, but recent observations of a few bursts with \textit{Fermi} GBM have confirmed previous indications
from BATSE of the 
presence of an underlying thermal component. 
Photospheric emission is indeed expected when the relativistic outflow emerging from the central engine becomes transparent to its own radiation, with a quasi-blackbody spectrum in absence of additional sub-photospheric dissipation. However, its intensity strongly depends on the acceleration mechanism --  thermal or magnetic -- of the flow.}         
{We aim to compute the thermal and non-thermal emissions 
produced by an outflow 
with a variable Lorentz factor, where the power ${\dot E}_{\rm iso}$  injected at the origin is partially thermal (fraction $\epsilon_{\rm th}\le 1$) and partially magnetic (fraction $1-\epsilon_{\rm th}$). 
The thermal emission is produced at the photosphere, and the non-thermal emission in the optically thin regime.
Apart from the value of $\epsilon_{\rm th}$,  we want 
to test how the other model parameters affect the observed ratio of the thermal to non-thermal emission. }
{We followed the adiabatic cooling of the flow from the origin to the photosphere and computed the emitted radiation, 
which is a sum of modified black bodies at different temperatures. 
If the non-thermal emission comes from internal shocks, 
it is obtained from a multi-shell model where a fraction of the energy dissipated in shell collision 
is transferred to electrons and radiated via the synchrotron mechanism. If, conversely, the non-thermal 
emission originates in magnetic reconnection, the lack of any detailed theory for this process forced us 
to use a very simple parametrisation to estimate the emitted spectrum.     }
{If the non-thermal emission is made by internal shocks, we self-consistently obtained the light curves and spectra of the thermal and non-thermal components for any distribution of the Lorentz
factor in the flow. If the non-thermal emission results from magnetic reconnection we were unable to produce a light curve and could only compare the respective non-thermal and thermal spectra.  In the different considered cases, we varied the model parameters to see when the thermal component in the light curve and/or spectrum is likely to show up or, on the contrary, to be
hidden. We finally compared our results to the proposed evidence for the presence of a thermal component in GRB spectra. Focussing on GRB 090902B and GRB 10072B, we showed how these observations can be used to constrain the nature and acceleration mechanism of GRB outflows.}
{}


\maketitle
\section{Introduction}
The first GRB spectra were obtained by the gamma and 
X-ray spectrometers on board the IMP-6 satellite and the Apollo 16 spacecraft
\citep{cline_1973, metzger_1974}. These early observations were complemented by 
the large sample of 143 spectra collected by the Konus 
experiments on the Venera probes from 1978 to 1980 \citep{mazets_1981}. It was shown that these spectra
could be fitted 
by a power law with an exponential cutoff or a broken power law \citep{cline_1975}, and various physical 
processes were invoked to explain this shape, such as optically-thin thermal bremsstrahlung \citep{gilman_1980},
Compton scattering of soft photons by non-thermal electrons \citep{zdziarski_1986} or synchrotron
emission by thermal or non-thermal electrons \citep{brainerd_1987}. Following the launch of 
the Compton Gamma-Ray Observatory, the Burst and Transient Source Experiment (BATSE) provided the first solid indications
that GRBs were located at cosmological distances and confirmed the broken power law 
shape of the spectra that was represented by the phenomenological Band function \citep{band_1993}.   
Possible cyclotron or annihilation lines found by previous experiments were not seen by BATSE.

In the context of the cosmological models that were developed hereafter, the spectra were 
generally interpreted 
in terms of synchrotron emission from shock accelerated electrons (see e.g. \citealt{piran_1999}) and therefore believed
to be mostly non-thermal. The possibility that a thermal contribution could also be present was, however, considered by \citet{meszaros_2002}, and was 
supported by several observational indications.
The first one came from the very hard low-energy spectral slopes that are
found in some BATSE bursts during at least part of the evolution \citep{preece_1998,ghirlanda_2003}. While the commonly observed value of the
low-energy spectral index is $\alpha\sim -1$, it reaches $0.5$ to $1$ in these events, suggesting the presence of
a Rayleigh-Jeans contribution.     
Then \citet{ryde_2004, ryde_2005} proposed to fit all GRB spectra in the BATSE range with the combination
of a thermal and a power law component. Using a time resolved analysis, he  
showed that during a pulse
the temperature first stays approximately constant before decaying as a power law of temporal index 
close to $-2/3$. 

In the previous examples the identified thermal component represented a major contribution
responsible for the peak of the $E^2N(E)$ spectrum.
A different result was obtained by \citet{guiriec_2011} who found a sub-dominant thermal component in 
the \textit{Fermi}-Gamma-ray Burst Monitor (GBM) spectrum of GRB 100724B, accounting for a few percents of the energy released by the burst. The evolution
of the temperature was not correlated to the peak energy of the non-thermal component, fitted by a Band function. 
Similar results have been found in other bright GBM bursts, such as GRB 110721A \citep{ryde_2012}, GRB 081207, and GRB 110920 \citep{mcglynn_2012}.
This improved characterisation of thermal components in the prompt emission of GRBs is allowed by the larger spectral coverage of GBM (8 keV--40 MeV) compared to BATSE (20 keV--2 MeV), leading to a better quality of the spectral fits. 

On the theoretical side, a thermal emission originating from the photosphere is a natural
prediction of most models based on the generic fireball scenario \citep{paczynski_1986,goodman_1986,shemi_1990,meszaros_1993}.  
Moreover, if the acceleration of the outflow has a thermal origin (as would be the case if it was powered at its basis 
by neutrino-antineutrino annihilation), this photospheric emission would be very bright, outshining the 
non-thermal emission produced by internal shocks in the 100 keV - 1 MeV spectral range \citep{daigne_2002}. 
One is therefore faced with the following
alternative:  either most of the emission we observe is indeed this thermal component, but it has been 
Comptonised to produce a power law tail in the spectrum at high energy (see e.g. \citealt{thompson_1994, rees_2005,giannios_2007, beloborodov_2010}) and complemented at low energy by additional processes (e.g. \citealt{peer_2006,vurm_2011}), or  the acceleration has 
a magnetic origin (see e.g. \citealt{begelman_1994, daigne_2002b, vlahakis_2003, komissarov_2009, tchekhovskoy_2010, komissarov_2010, granot_2011}) and the fraction 
$\epsilon_{\rm th}$ of thermal energy in the flow is much smaller than unity.

In this work we explore  the consequences of this second possibility. We self-consistently compute the photospheric
thermal emission and the non-thermal emission from internal shocks. 
If the internal dissipation is dominated by magnetic reconnection rather than internal shocks, we get the non-thermal spectrum in a very simple parametrised way. We discuss the conditions for the thermal emission 
to show up or, on the contrary, to be hidden. The paper is organised as follows: in Sect.~2 we describe the geometry and
thermodynamics of the flow and explain our method to compute the photospheric and non-thermal emissions; our results are presented
in Sect.~3 and discussed in Sect.~4. Finally, Sect.~5 is the conclusion.    

\section{Model description}
\subsection{Geometry and thermodynamics of the flow}
\begin{figure}
\begin{center}
\begin{tabular}{c}
\includegraphics[width=0.4\textwidth]{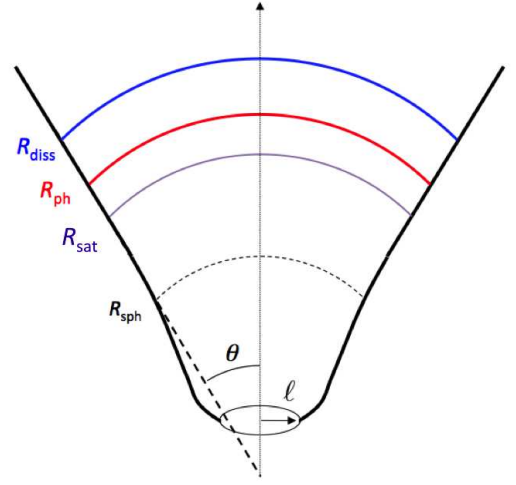} 
\end{tabular}
\end{center}

\caption{
{\bf Schematic view of the problem geometry.} The flow emerges from the central engine through a ``circular opening''
of radius $\ell$. Beyond a radius $R_{\rm sph}$ it expands radially within a cone of half opening $\theta$. The acceleration
is completed at $R_{\rm sat}$. The photosphere is located at $R_{\rm ph}$ and dissipation of kinetic and/or magnetic energy
takes place at $R_{\rm diss}$.
}
\label{schema}
\end{figure}

We consider a schematic model where the flow that emerges from the central engine is accelerated by the conversion of thermal and/or magnetic energy to kinetic energy. 
We do not specify the initial geometry of the flow
(which can be largely governed by magnetic forces), but we assume that beyond a radius $R_{\rm sph}$ it 
becomes spherically symmetric within a cone of half opening angle $\theta$. 
We also define the radius $R_\mathrm{sat}$, where the acceleration is essentially complete and suppose that $R_{\rm sph}<R_\mathrm{sat}$.     
The total injected power in the flow is ${\dot E}$, with a fraction $\epsilon_{\rm th}$ 
in thermal form. The temperature $T_0$ at the origin of the flow can then be obtained from 
\begin{equation}
\label{eqn_eth}
{\dot E}_{\rm th}=\epsilon_{\rm th}\,{\dot E}=\epsilon_{\rm th}\,{\Omega\over 4\pi}\,{\dot E}_{\rm iso}
=a T_0^4\,c\times S_0\, ,
\end{equation}
where ${\dot E}_{\rm iso}$ and $S_0=\pi {\ell}^2$ are, respectively, the isotropic injected power and the section of the flow at the
origin (see \fig{schema}). The fraction of solid angle is ${\Omega\over 4\pi}\simeq {\theta^2\over 4}$ 
(we count only one jet to be consistent with the definition of $S_0$)
and $a$ is the radiation constant. We finally get
\begin{equation}
\label{eqn_t0}
T_0\simeq 0.66\,\epsilon_{\rm th}^{1/4}\,\theta_{-1}^{1/2}\,{\dot E}_{{\rm iso},53}^{1/4}\,{\ell}_7^{-1/2}\ \ {\rm MeV}\, ,
\end{equation}
with the opening angle, injected power, and radius of the jet in units of 0.1 rad, $10^{53}$ erg.s$^{-1}$, 
and $10^7$ cm, respectively .   

We obtain the flow equations assuming that no dissipation takes places below the photosphere (so that
the emerging spectrum at transparency will be thermal only).   
Mass and entropy conservation then lead to
\begin{eqnarray}
\label{eqn_cons_mass}
\beta \Gamma\,\rho\,S=Cst\, ,\\
\label{eqn_adia}
{T\over \rho^{1/3}}=Cst \, ,
\end{eqnarray}
where $\rho$ is the comoving density, $S$ the surface perpendicular to the flow, $\beta=v/c$,
and $\Gamma=(1-\beta^2)^{-1/2}$. From Eqs.~\ref{eqn_cons_mass}-\ref{eqn_adia} we get
\begin{equation}
\label{eqn_adia_mass}
\beta \,\Gamma\,T^3\,S=Cst\ .
\end{equation}
Using \eq{eqn_adia_mass} we can obtain the temperature at any radius $R>R_{\rm sph}$ even if we ignore
the details of the geometry from the basis of the flow up to $R_{\rm sph}$. 
With $S(R)=\pi\,\theta^2\,R^2$ and assuming that $\beta \sim 1$ already close to the origin
we have
\begin{equation}
T(R)\simeq T_0\times \left(\theta^{-2/3}R^{-2/3}{\ell}^{2/3}\,\Gamma^{\,-1/3}\right)
\end{equation}
so that, in the observer frame
\begin{equation}
\label{eqn_tobs}
T_{\rm obs}(R)={\Gamma\,T(R)\over 1+z}\simeq {T_0\over 1+z}\times 
\left(\theta^{-2/3}R^{-2/3}{\ell}^{2/3}\,\Gamma^{2/3}\right)\, ,
\end{equation}
where $z$ is the burst redshift.

At the photospheric radius $R_{\rm ph}$ (supposed to lie beyond $R_{\rm sph}$) the thermal luminosity is given by
\begin{equation}
\label{eqn_lth}
L_{\rm th}=\Gamma_{\rm ph}^2\ a\,T^4(R_{\rm ph})\,c\times S(R_{\rm ph})={\dot E}_{\rm th}\times 
\left(\theta^{-2/3}R_{\rm ph}^{-2/3}{\ell}^{2/3}\,\Gamma_{\rm ph}^{2/3}\right)\, .
\end{equation}
Equations~\ref{eqn_tobs} and \ref{eqn_lth} correspond to the usual scaling of the fireball scenario \citep{meszaros_2000,meszaros_2002,daigne_2002} with, however, a modified normalization that takes into account both the geometry and the mixed energy content of the outflow. Conversely, including sub-photospheric dissipation as 
in \citet{giannios_2012} would change the scaling.

To estimate the photospheric radius we assume that most of the acceleration is completed
at $R_{\rm ph}$\footnote{See Appendix A for a short discussion of the case where $R_\mathrm{sat}>R_{\rm ph}$.}. 
This is, for example, the case in the simulations made by \citet{tchekhovskoy_2010},
where the Lorentz factor sharply increases beyond the stellar radius, when the flow suddenly becomes unconfined.
Then, in a first approximation, the photospheric radius of a given shell
writes (e.g. \citealt{piran_1999, meszaros_2000, daigne_2002})
\begin{equation}
\label{eqn_rph}
R_{\rm ph} \simeq
{\kappa\,{\dot M}\over 8\pi\,c\,\Gamma^2}
\simeq
2.9 \times10^{13}\,{\kappa_{0.2}\,{\dot E}_{\rm iso,53}\over
(1+\sigma)\,\Gamma_2^3}\ \ {\rm cm}\, ,
\end{equation}
where $\kappa$ ($\kappa_{0.2}$ in units of 0.2 cm$^2$.g$^{-1}$)
is the material opacity and $\Gamma$ ($\Gamma_2$ in units of 100) the Lorentz factor of the shell.
The flow keeps a magnetisation $\sigma$ at the end of acceleration so that 
${\dot E}/(1+\sigma)$ is the injected kinetic power ${\dot E}_{\rm K}$. 
In the case of a passive magnetic field that is carried by the outflow without contributing to its acceleration \citep{spruit_2001}, the magnetisation $\sigma$ equals 
\begin{equation}
\sigma_{\rm passive} = (1-\epsilon_{\rm th})/\epsilon_{\rm th}\, ,
\label{eq:signa_passive}
\end{equation}
 corresponding to a pure and complete thermal acceleration. Efficient magnetic acceleration leads to $\sigma < \sigma_\mathrm{passive}$, whereas $\sigma > \sigma_\mathrm{passive}$ corresponds to an inefficient magnetic acceleration, for instance with no conversion of magnetic into kinetic energy and some conversion of thermal into magnetic energy. 

When the shell reaches the photospheric radius, it releases its thermal energy
content while a fraction of the remaining energy (kinetic or magnetic) can be
dissipated farther away
 at a radius $R_{\rm diss}$ by internal shocks for $\sigma \la 0.1-1$, or reconnection for higher magnetisation, 
  contributing to
the non-thermal emission of the burst.    
It is therefore expected, on theoretical grounds, that thermal and non-thermal components both contribute to the observed emission \citep{meszaros_2002,daigne_2002}. As mentioned in the introduction, this was already supported by BATSE results,  with new evidence now coming from  \textit{Fermi} \citep{guiriec_2011,zhang_bb_2011}.
In Sect.~\ref{sec:results}
we present  synthetic bursts showing both contributions. We explain below our method to compute the thermal and non-thermal emission.

\subsection{Thermal emission}
\label{sec_calc_ph}
The thermal emission can be computed from Eqs.~\ref{eqn_eth}, \ref{eqn_t0}, \ref{eqn_tobs}, \ref{eqn_lth}, and \ref{eqn_rph}, for a given set 
of central engine parameters
$\epsilon_{\rm th}$, $\sigma$, $\ell$, $\theta$, ${\dot E}_{\rm iso}$, and a distribution of 
the Lorentz factor in the flow. Both the thermal luminosity and observed temperature are related to
the injected power and temperature at the origin of the flow via the same factor 
\begin{equation}
\label{eqn_phi}
\Phi=\left(\theta^{-2/3}R_{\rm ph}^{-2/3}{\ell}^{2/3}\,\Gamma_{\rm ph}^{2/3}\right)\propto {\dot E_{\rm iso}}^{-2/3}\
\Gamma_{\rm ph}^{8/3}\, .
\end{equation}
If a constant $\dot E_{\rm iso}$ is assumed, the luminosity and temperature directly
trace the distribution of the Lorentz factor $\Gamma(s)$, 
where the Lagrangian coordinate $s$ is the distance to the front of the flow at the end of the acceleration stage ($s/c$ is the ejection time of the shell).
The expanding flow becomes progressively transparent (starting from the front) and the
contribution of
a shell located at $s$ is approximately received at an observer time 
$t_{\rm obs}=(1+z)s/c$ \citep{daigne_2002}.  
However, since the different parts of
the flow do not become transparent at the same radius (because $R_{\rm ph}\propto \Gamma^{-3}$), 
additional differences in arrival time of the order of
\begin{equation}
\label{dtobs_ph}
\Delta t_{\rm obs}=(1+z)\,{R_{\rm ph}\over 2\,c\,\Gamma^2}\simeq 49\,(1+z)\,{\kappa_{0.2}\,{\dot E}_{{\rm iso},53}\over
(1+\sigma)\,\Gamma_2^5}\ \ {\rm ms}
\end{equation}
should be included. They are, however, negligible as long as $\Delta t_{\rm obs}<t_{\rm var}$, the typical
variability time scale of the Lorentz factor.

The value of $\Delta t_{\rm obs}$ also gives the time scale of the luminosity decline 
after the last shell of the flow (emitted by the source at a time $\tau$) has reached the transparency radius (high-latitude emission). 
Unless the Lorentz 
factor of this shell is small or the burst has a very short duration,
the drop in luminosity for $t_{\rm obs}>(1+z)\tau$ is very steep, having initially a temporal decay index \citep[see e.g. Sect.~6 in][]{beloborodov_2011}
\begin{equation}
\alpha={d\,{\rm Log}\,L_{\rm th}\over d\,{\rm Log}\,t_{\rm obs}}\simeq 2{(1+z)\tau \over \Delta t_{\rm obs}}\gg 1\ .
\end{equation}
This shows that in models where the prompt emission comes from a Comptonised photosphere, the early  
decay of index $\alpha\sim 3 - 5$ observed in X-rays cannot be explained by the high latitude emission and should instead be related to
an effective decline of the central engine \citep{hascoet_2012}. 

The expected count rate in a given spectral range and the resulting spectrum are 
obtained from the luminosity and temperature evolution (Eqs.~\ref{eqn_tobs} and \ref{eqn_lth}). 
However, we do not use a true Planck function for the elementary spectrum
corresponding to a given temperature. As discussed in \citet{goodman_1986} and \citet{beloborodov_2010}, geometrical 
effects at the photosphere lead to a low-energy spectral index close to $\alpha=+0.4$ instead of
$\alpha=+1$ for a Raleigh-Jeans spectrum (see also \citealt{peer_2008}). We therefore adopt a ``modified
Planck function'' having the modified spectral slope at low energy, an exponential cutoff
at high energy, peaking at $\simeq 3.9 \times k T$ as a Planck function in $\nu F_{\nu}$, and carrying the same total energy.

\begin{figure*}
\begin{center}
\begin{tabular}{cc}
\includegraphics[width=0.45\textwidth]{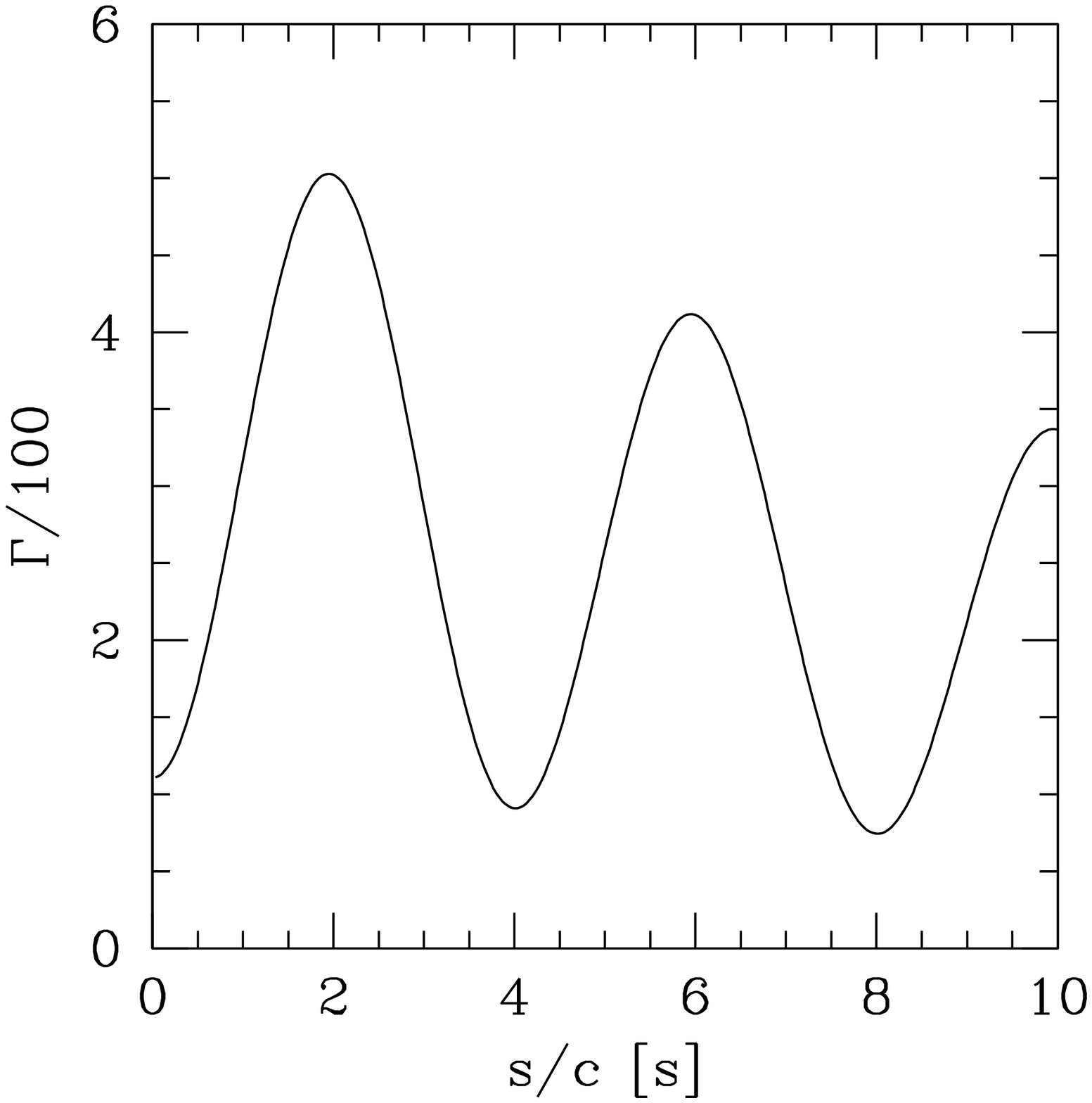} & \includegraphics[width=0.45\textwidth]{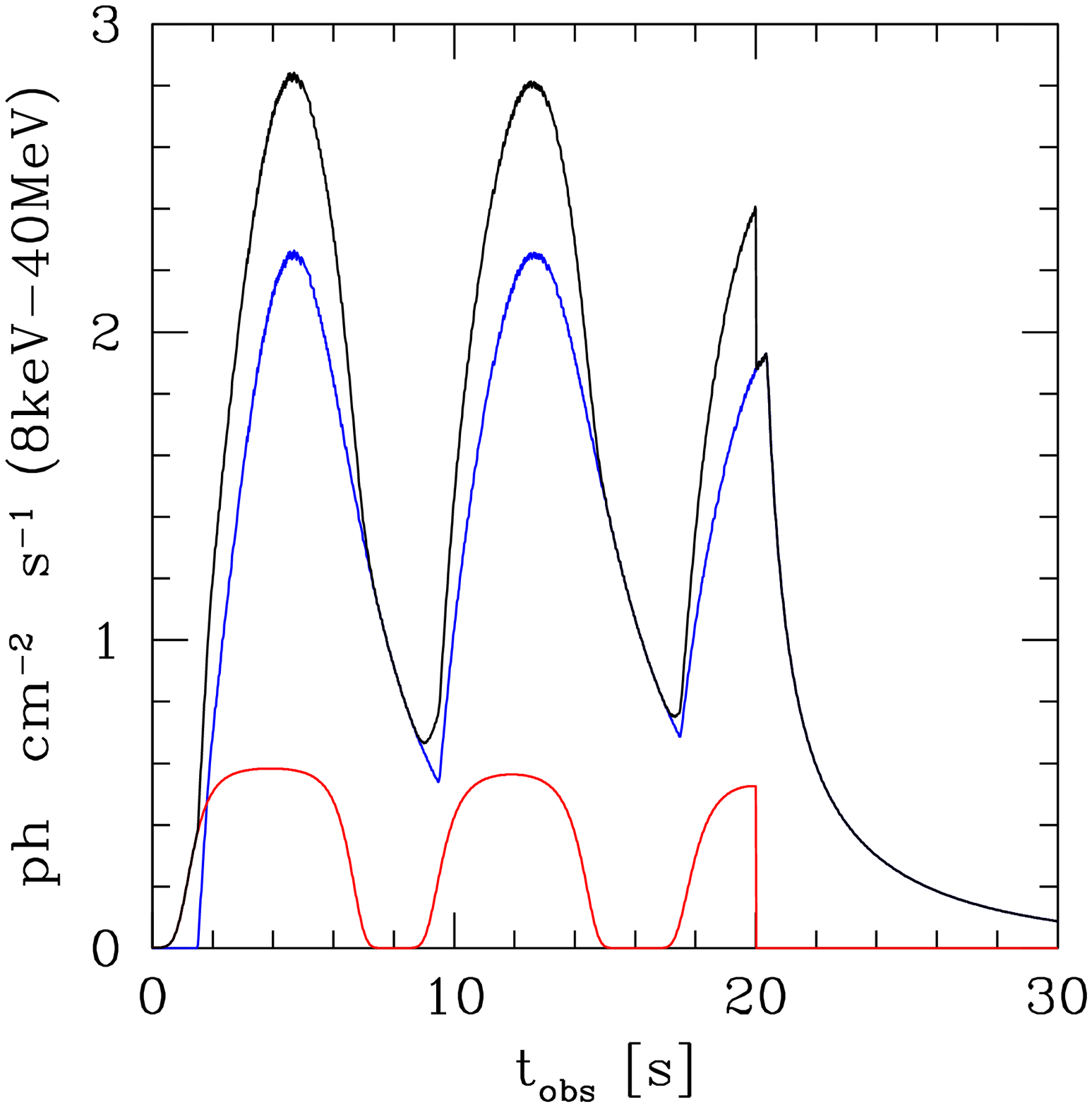} \\
\includegraphics[width=0.45\textwidth]{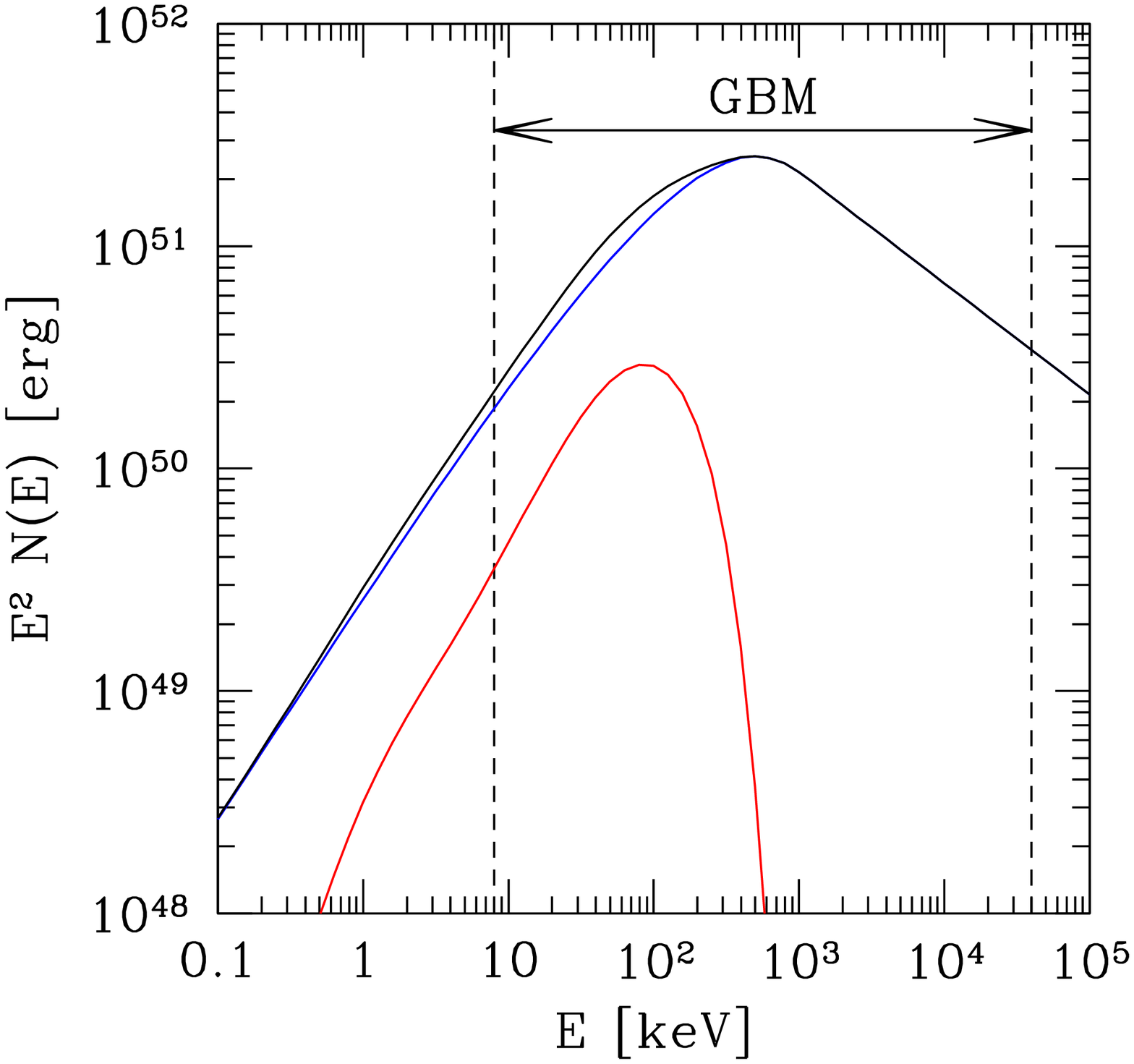} & \includegraphics[width=0.45\textwidth]{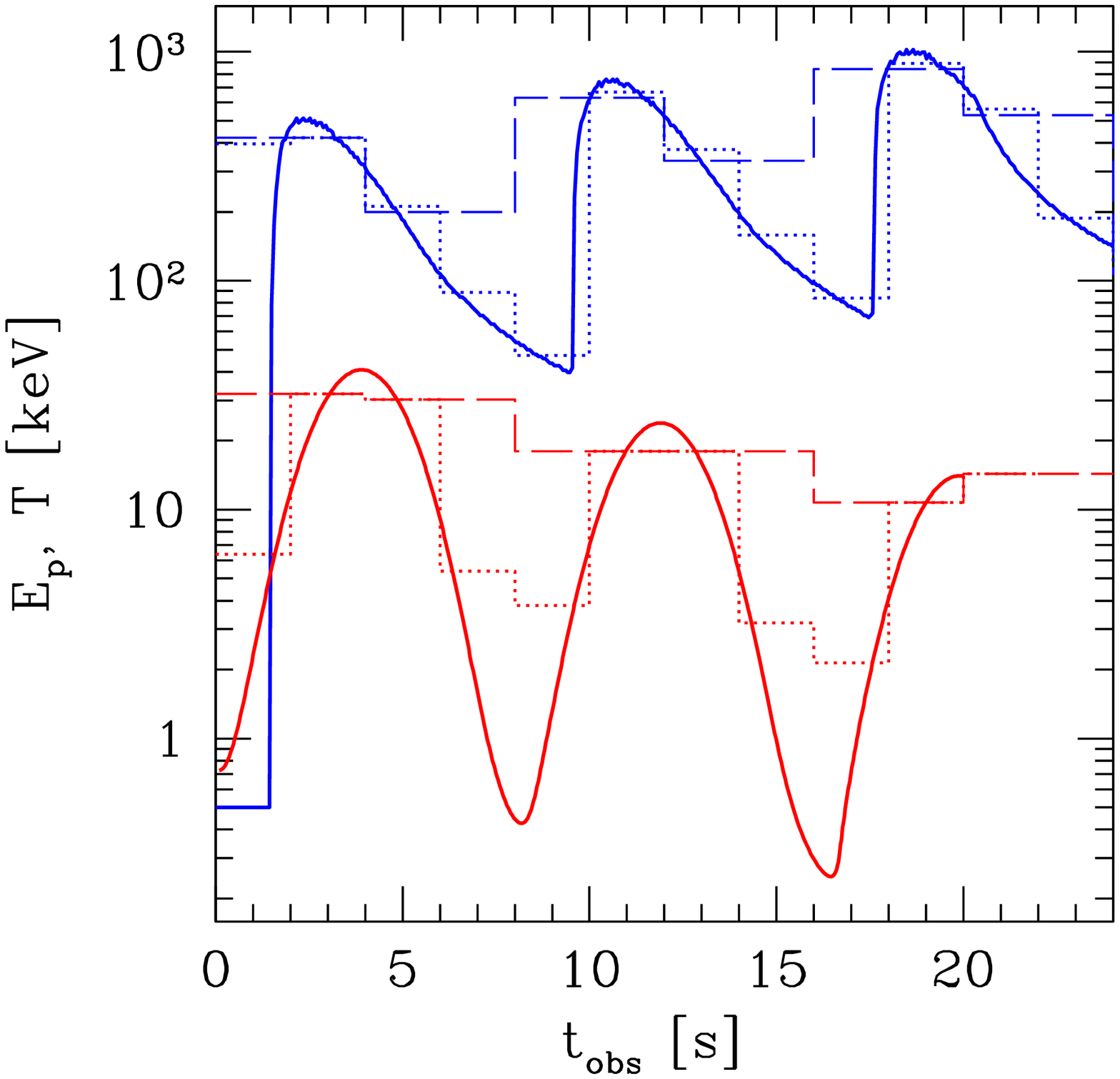}
\end{tabular}
\end{center}

\caption{
\textbf{Thermal and non-thermal emission from a variable outflow -- internal shock framework.} 
\textit{Top left: } initial distribution of the Lorentz factor in the flow. 
\textit{Top right: } thermal (red), non-thermal (blue), and total (black) photon flux in the 8 keV - 40 MeV spectral range.
\textit{Bottom left:} thermal (red), non-thermal (blue), and total (black) time-integrated spectra. 
\textit{Bottom right}: instant temperature (red) of the photospheric emission and instant peak energy (blue) of the internal shock 
emission. 
The dotted and dashed lines correspond to the temperature and peak energy averaged over 
time intervals of 2 and 4 s, respectively.  
The adopted flow parameters are
${\dot E}_{\rm iso}=10^{53}$ erg.s$^{-1}$, $\epsilon_{\rm th}= 0.03$, $\sigma=0.1$, $\ell=3\,10^6$ cm, and $\theta=0.1$ rad;
a redshift $z=1$ is assumed.}
\label{fig_variable}
\end{figure*}

\subsection{Non-thermal emission}
\label{sec:nonthermal}
We first estimate the non-thermal emission assuming that it comes from internal shocks \citep{rees_1994}. For a given distribution
of the Lorentz factor, we obtain light curves and spectra 
using the simplified model of \citet{daigne_1998}, where the outflow
is represented by a large number of shells that interact by direct collisions only (see also \citealt{kobayashi_1997}). 
The elementary spectrum for each collision is a broken power law with the break at the synchrotron energy. 
The adopted values for the two spectral indices at low and high energy respectively are $\alpha=-1$ and $\beta=-2.25$. The expected 
value for $\alpha$ in the fast cooling regime should normally be $-1.5$ (see e.g. \citealt{sari_1998, ghisellini_2000}), but detailed radiative models including 
inverse Compton 
scattering
in Klein-Nishina regime tend to produce harder $\alpha$ slopes \citep{derishev_2001, bosnjak_2009, nakar_2009, daigne_2011}, close to the typical observed value $\alpha \simeq -1$ \citep{preece_2000, kaneko_2006, nava_2011, goldstein_2012}. 
The global efficiency of internal shocks is given by the product 
\begin{equation}
\label{eqn_efficiency_is}
f_{\rm IS}=\epsilon_e\times f_{\rm diss}\, ,
\end{equation}
where $f_{\rm diss}$ is the efficiency for the dissipation of energy in shocks and $\epsilon_e$ the fraction of the
dissipated energy transferred to electrons and eventually radiated. Typical values of $f_{\rm IS}$ do not exceed a few percents.
We used this approach to compute non-thermal light curves and spectra in Sect.~3.1 for the case of a low magnetisation  $\sigma\la 0.1-1$.

The presence of magnetic fields reduces shock efficiency and may even prevent 
shock formation for $\sigma \ga 1$ \citep{mimica_2010, narayan_2011}. 
Then for $\sigma \ga 1$, the magnetic field cannot be ignored, and energy must be extracted by magnetic reconnection (e.g. \citealt{thompson_1994, spruit_2001}),
possibly triggered by internal shocks \citep{zhang_2011}. 
Our limited understanding of the relevant processes
does not allow a reliable description of the resulting emission. Considering these difficulties,
we have adopted in Sect.~3.2 a very basic and simple point of view. We do not try to predict the burst profile and
obtain the spectrum in the following way: we suppose that a fraction $f_{\rm \,Nth}$ of the total 
injected energy eventually goes into non-thermal emission with a spectrum represented by a Band
function with low- and high-energy spectral indices $\alpha=-1$ and $\beta=-2.25$, and a peak energy obtained from the 
Amati relation \citep{amati_2002}
\begin{equation}
\label{eqn_amati}
E_{\rm p}\simeq 130\,\left[f_{\rm \,Nth}\,{E}_{\rm iso}\over 10^{52}\ {\rm erg}\right]^{0.55}\ \ {\rm keV}\ ,
\end{equation}
where the exponent and normalisation values are taken from \citet{nava_2012}.
The validity of the Amati relation is strongly debated (see e.g. \citealt{nakar_2005,band_2005,kocevski_2012,collazzi_2012, ghirlanda_2012}). It is not
clear if it corresponds to an intrinsic property of gamma-ray bursts or if the relation results from a complex chain 
of selection effects (threshold for burst detection, and various conditions for the measure of the redshift and 
peak energy). For the purpose of the present study we do not address this issue and use  
\eq{eqn_amati} simply because it is approximately satisfied 
by the sample of long bursts for which the peak energy and isotropic radiated energy have been measured.        

\begin{figure*}
\begin{center}
\begin{tabular}{ccc}
\includegraphics[width=0.3\textwidth]{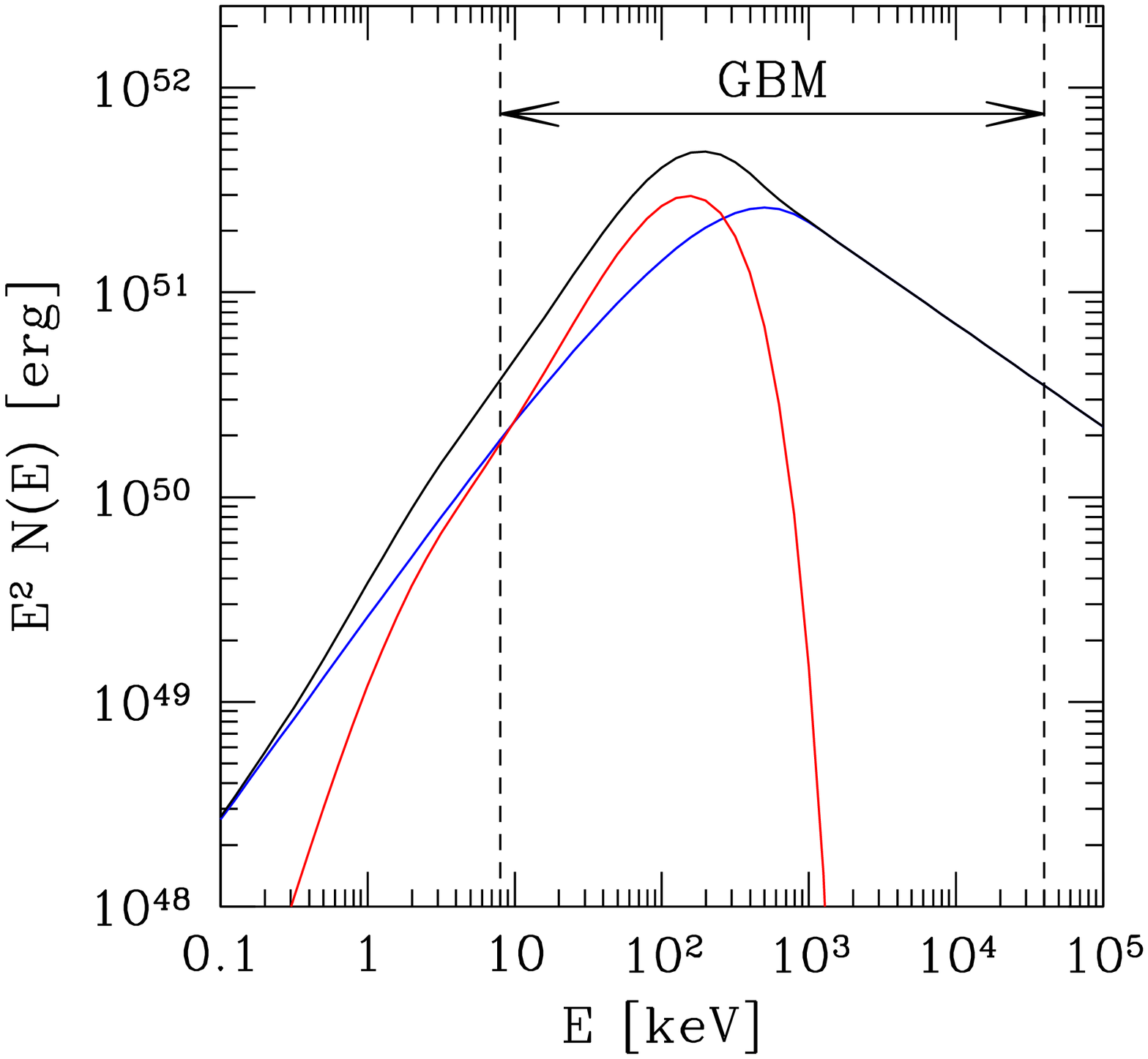} &
\includegraphics[width=0.3\textwidth]{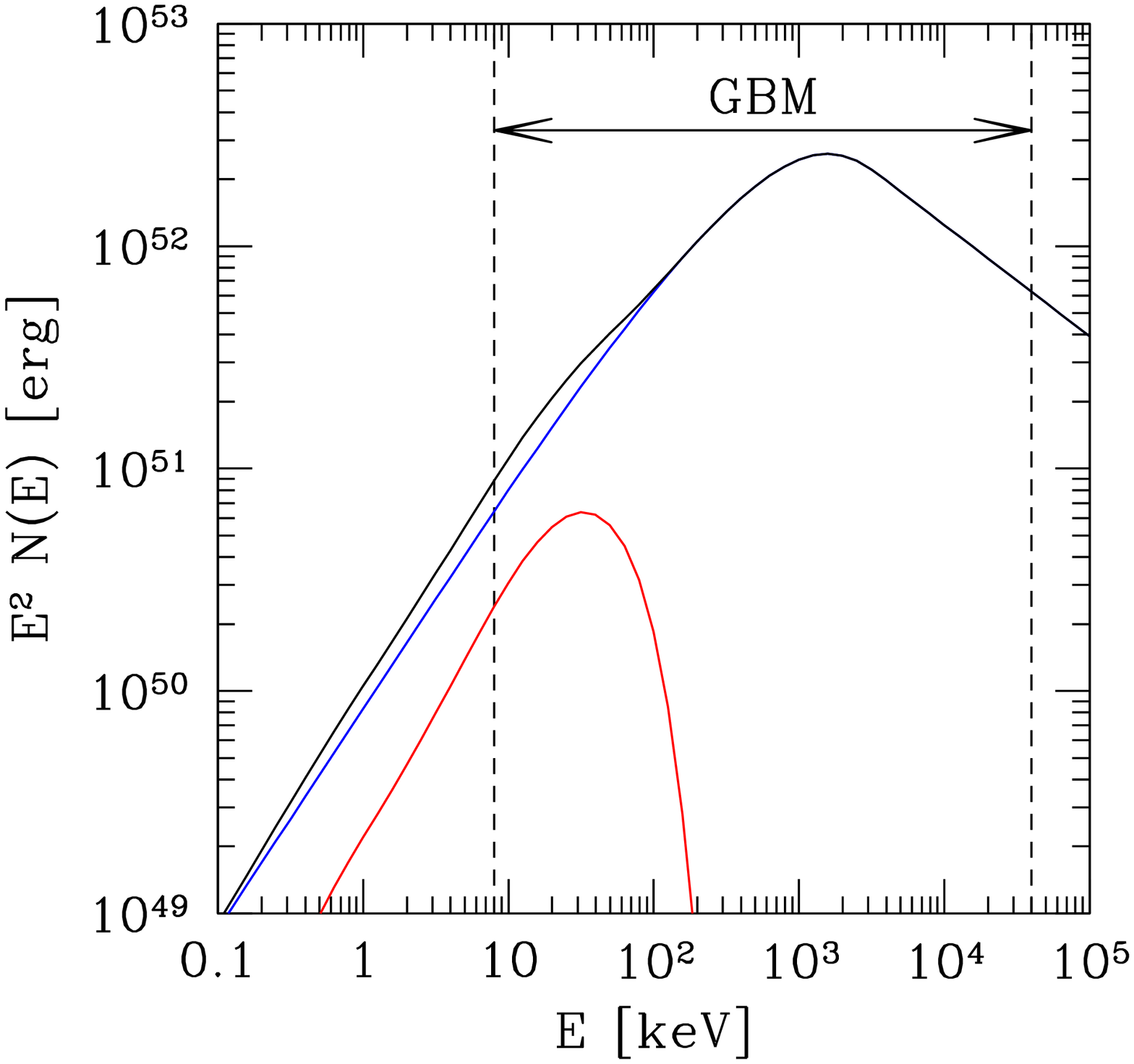}  &
\includegraphics[width=0.3\textwidth]{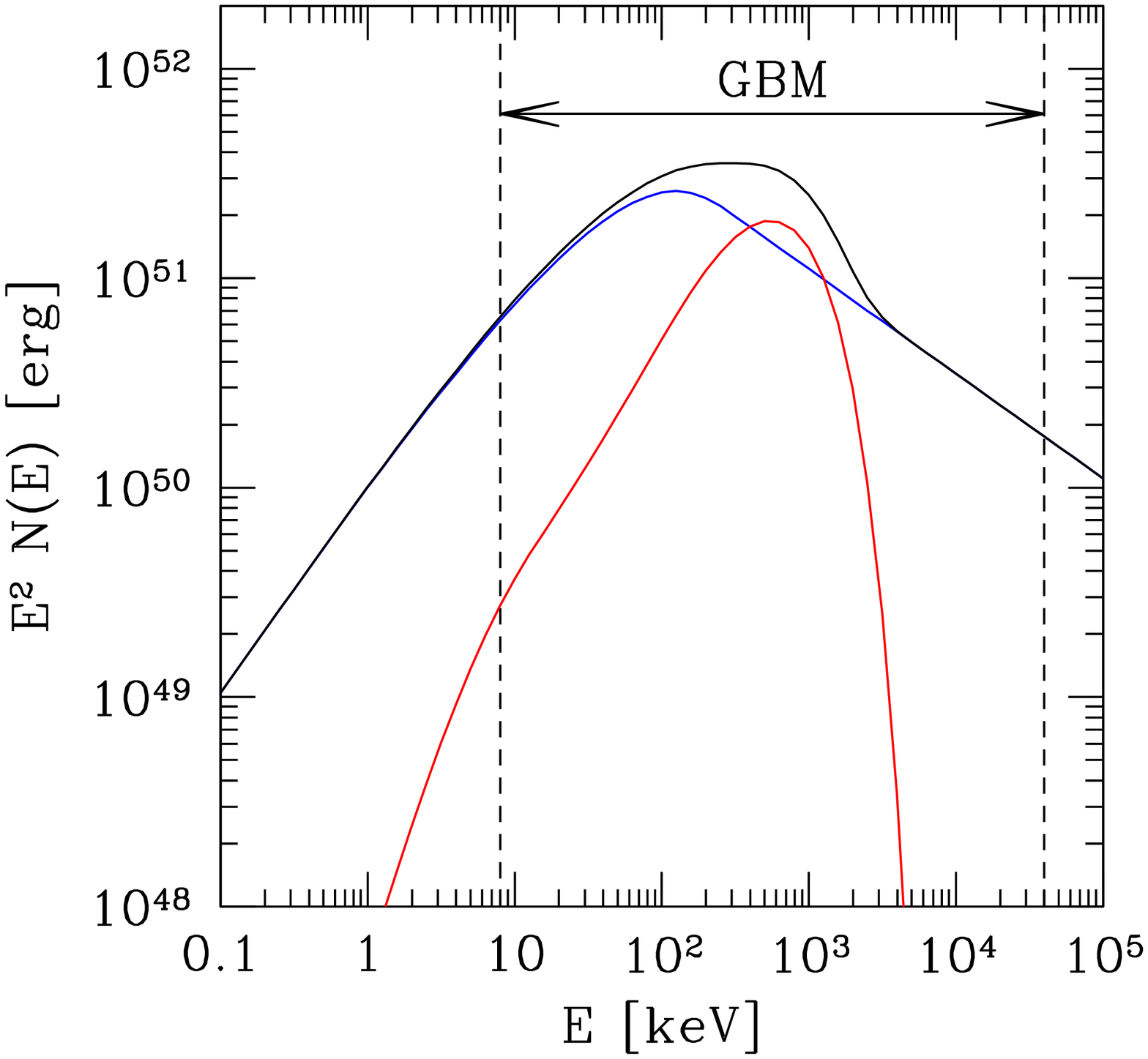} \\
\includegraphics[width=0.3\textwidth]{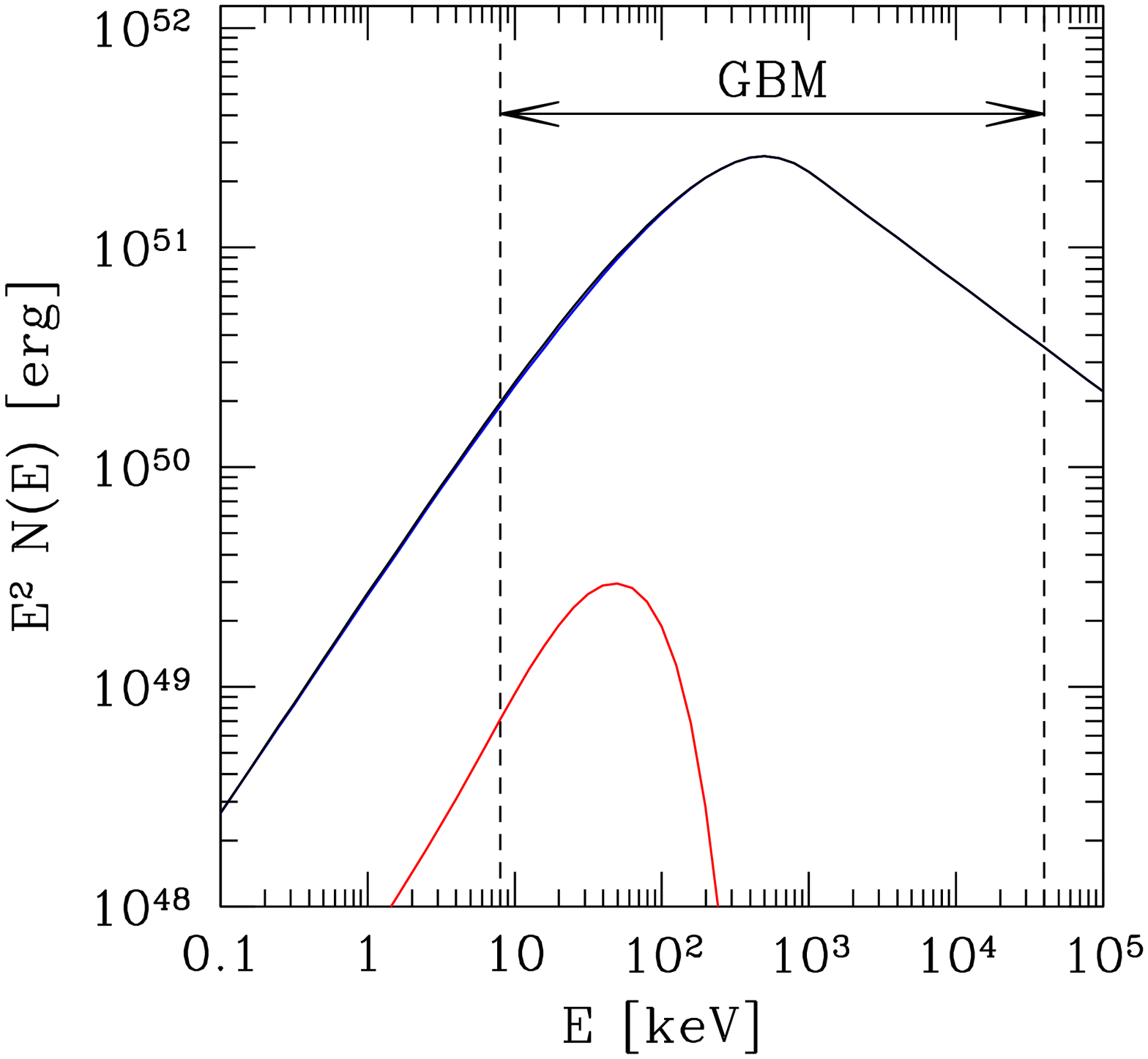} &
\includegraphics[width=0.3\textwidth]{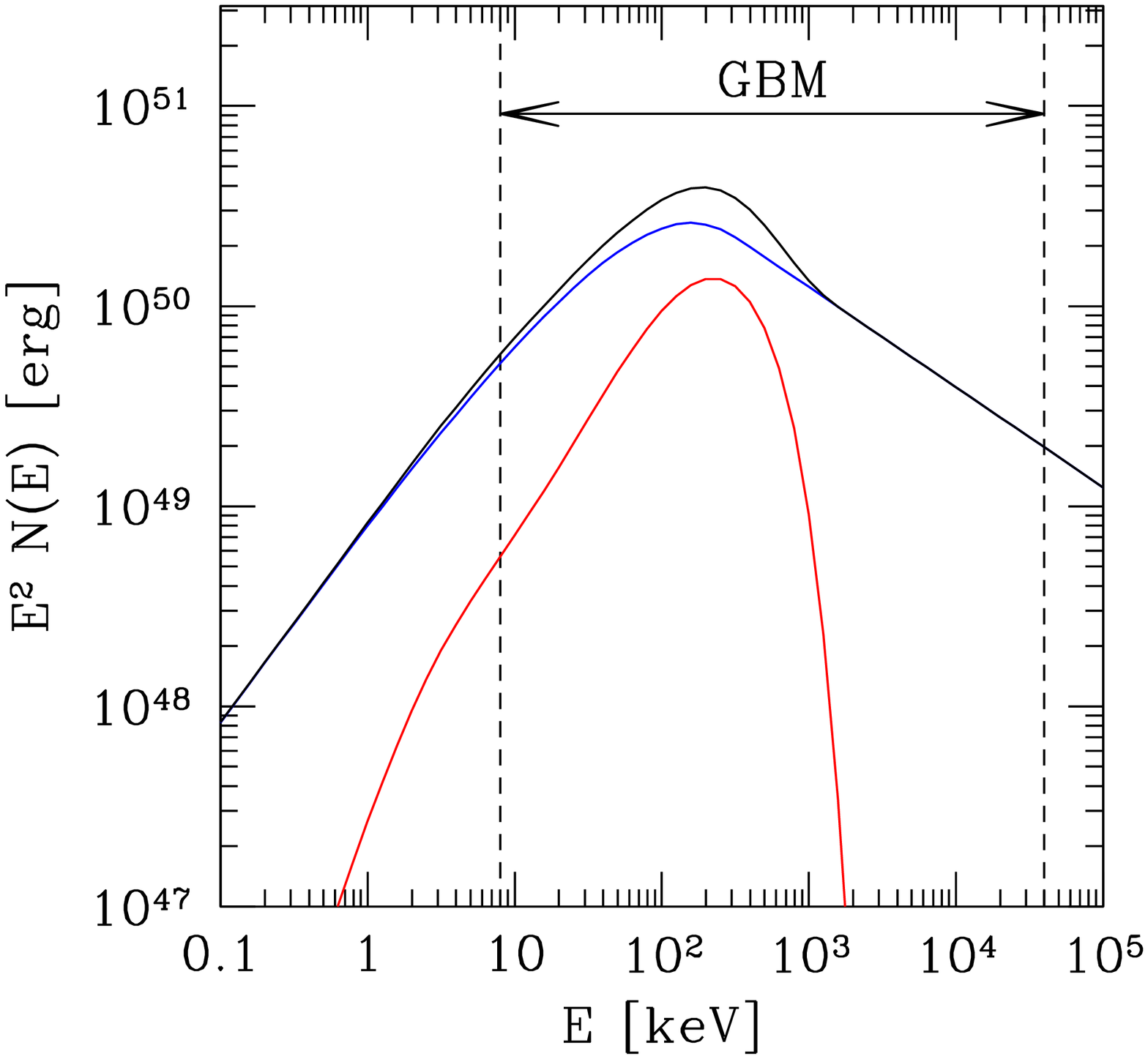}  &
\includegraphics[width=0.3\textwidth]{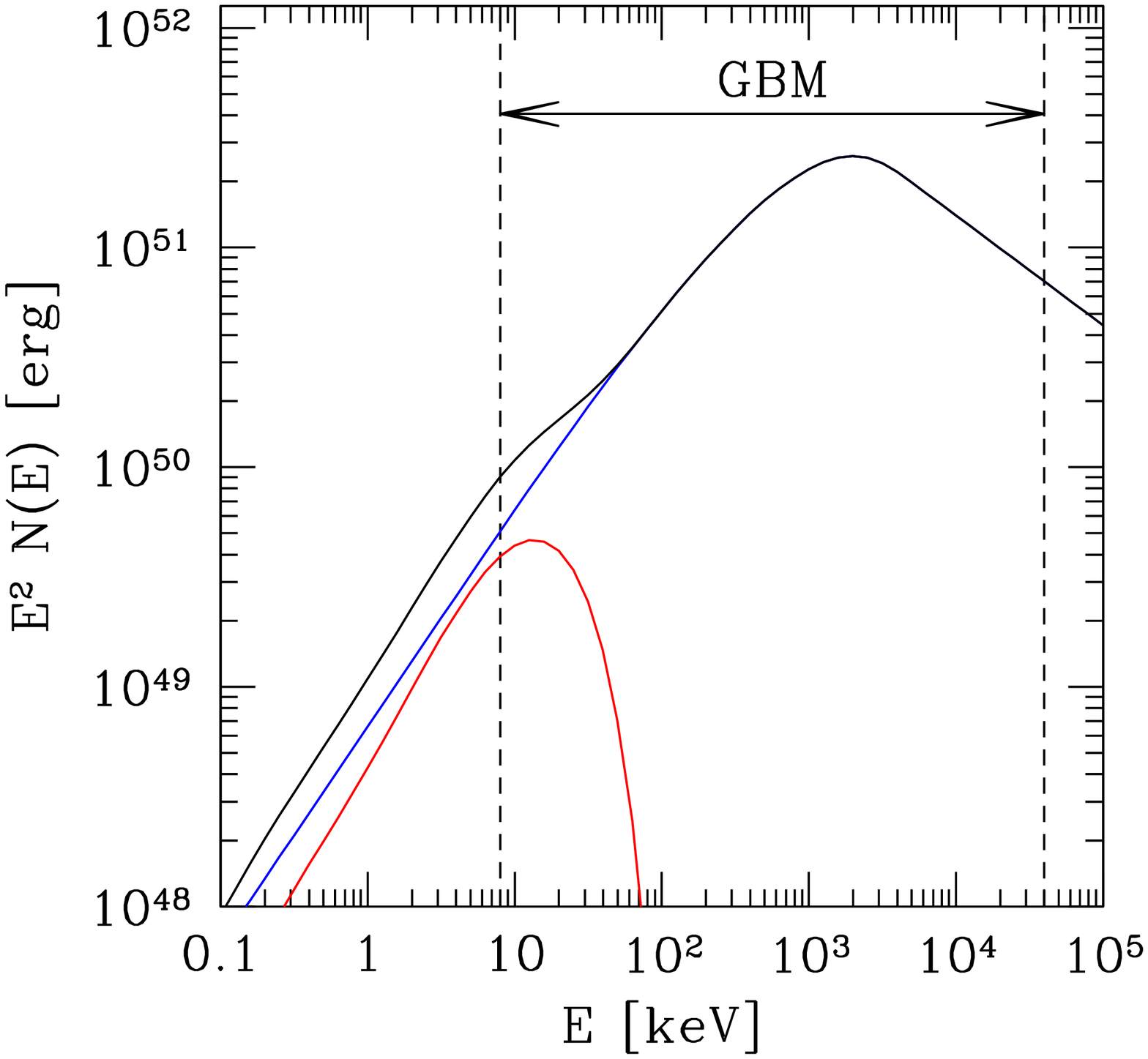}
\end{tabular}
\end{center}

\caption{
\textbf{Thermal and non-thermal emission from a variable outflow -- 
internal shock framework: impact of the outflow parameters.} 
Starting from the reference case shown in \fig{fig_variable} (with a redshift $z=1$) we first change
the thermal fraction $\epsilon_{\rm th}$ to $0.3$ (top left panel) and $0.003$ (bottom left); then 
the isotropic power $\dot{E}_{\rm iso}$ to  $10^{54} \ \mathrm{erg \ s^{-1}}$ (top middle) and 
$10^{52} \ \mathrm{erg \ s^{-1}}$ (bottom middle); and finally
the Lorentz factor is multiplied (top right) and divided (bottom right) by a factor of 2.
Each time we vary a parameter, all the others keep the value corresponding to the reference case. 
}
\label{fig_influence_param}
\end{figure*}

\section{Results}
\label{sec:results}
\subsection{Non-thermal emission from internal shocks}
To study the relative intensities of the thermal and non-thermal emission components  
we have considered as an example a flow ejected by a central source active for a duration $s_{\rm tot}/c=10$ s, where the Lorentz factor takes the form 
\begin{equation}
\label{eqn_gp_complex}
\Gamma(s)=333 \left\lbrace 1+\frac{2}{3} \ {\rm cos}\left[5\,\pi \left( 1-{s\over s_{\rm tot}}\right)\right]\,\right\rbrace\times \exp{\left(-{s\over 2 s_{\rm tot}}\right)}\ .
\end{equation}
This distribution is arbitrary (see other possible examples in \citealt{daigne_1998}) and was adopted simply to produce a burst made of three pulses, i.e.
not too simple and not too complex. It is   
is shown in \fig{fig_variable} (top left panel) together with the non-thermal light curve (between $8$ keV and 40 MeV; top right panel)  
resulting from internal shocks. The related 
photospheric emission is shown in the same energy range for $\epsilon_{\rm th}=0.03$. We adopt a constant ${\dot E}_{\rm iso}=10^{53}$
erg.s$^{-1}$, $\sigma=0.1$, $\ell=3\,10^6$ cm and $\theta=0.1$ rad (i.e. $l/\theta = 300$ km). We also represent in the bottom left and bottom right panels the 
spectrum (thermal, non-thermal, and global) and the temporal evolution of the instantaneous peak energy and temperature of 
the non-thermal and thermal emissions. 
As the thermal spectrum is a superposition of elementary modified Planck functions at different temperatures,
its average spectral slope 
$\alpha$ (with $N(E)\propto E^{\alpha}$)
below the peak is close to $-1$. The asymptotic value $\alpha=+0.4$ is recovered only below a few keV, which corresponds
to the photospheric contribution with the lowest temperature.

It can be seen in \fig{fig_variable} (top right panel) that the emission is initially only
thermal as it takes a time 
\begin{equation}
\label{eqn_shiftb}
\frac{\Delta t_{\rm IS, 0}^{\rm obs}}{1+z}\simeq \left({R\over 2c\,\Gamma^2}\right)_0\simeq 0.42\ \ {\rm s} 
\end{equation}
for the first signal from internal shocks to arrive at the observer. In \eq{eqn_shiftb} the subscript ``0'' refers to the 
radius and Lorentz factor of the first shocked shell that contributes to the non-thermal emission.
At late times the situation is just the opposite: the photospheric emission abruptly stops at $t_{\rm obs}=20$ s,
while the emission from late internal shocks (both on- and off-axis) still contribute for about 10 s (observer frame).
The spectrum (\fig{fig_variable}, bottom left) is the sum of the thermal and non-thermal contributions. They  peak
at $200/(1+z)$ and $1000/(1+z)$ keV, respectively. 
Finally the plot of the temperature and of the
peak energy of the non-thermal spectrum as a function of observer time shows (\fig{fig_variable}, bottom right panel) that the former is more sensitive than the latter 
to the fluctuations of
the Lorentz factor (since $T_{\rm obs}\propto \Gamma^{8/3}$).

We now check how these results change when we vary the model parameters. We divide these
parameters into three groups describing respectively the geometry ($\theta$, $l$); 
the energetics and flow-acceleration mechanism (${\dot E}_{\rm iso}$
$\epsilon_{\rm th}$, $\sigma$); and  the ejecta structure ($s_{\rm tot}$, ${\cal C}$, ${\bar \Gamma}$),
where ${\cal C}$ and $\bar \Gamma$ are  the contrast (${\cal C}=\Gamma_{\rm max}/\Gamma_{\rm min}$)
and average of the Lorentz factor distribution. 

The geometry only affects the thermal emission (for a fixed ${\dot E}_{\rm iso}$). The dependence of $T_{\rm obs}$
and $L_{\rm th, \,iso}$ on $\theta$ and $\ell$ is weak. From Eqs.~\ref{eqn_t0}, \ref{eqn_tobs}, and \ref{eqn_lth} we get $T_{\rm obs}\propto \theta^{-1/6}\ell^{1/6}$
and $L_{\rm th, \,iso}\propto \theta^{-2/3}\ell^{2/3}$. Similarly, changing the magnetisation has a moderate effect
on the previous results, as long as $\sigma<1$. As explained above, for $\sigma\ga 1$ 
the whole theoretical framework adopted to compute the non-thermal emission probably becomes invalid. 

The consequence of increasing or decreasing the thermal fraction $\epsilon_{\rm th}$ is
illustrated in \fig{fig_influence_param} (left panels).  With $\epsilon_{\rm th}=0.3$ the thermal spectrum overtakes the
non-thermal spectrum between $20/(1+z)$ and $500/(1+z)$ keV, and the thermal component                    
represents about one third of the total in the light curve. 
Conversely, with $\epsilon_{\rm th}=0.003$ the contribution of the thermal component to the global spectrum
and the light curve is barely visible.  

The dependence of the temperature and thermal luminosity on the injected isotropic power ${\dot E}_{\rm iso}$
and average Lorentz factor ${\bar \Gamma}$
can be also obtained from Eqs.~\ref{eqn_t0}, \ref{eqn_tobs}, and \ref{eqn_lth} yielding
\begin{equation}
\label{eqn_tobs_depend}
T_{\rm obs} 
\propto
{\dot E}_{\rm iso}^{-5/12}\,{\bar \Gamma}^{8/3}\, ,
\end{equation}
\begin{equation}
\label{eqn_lth_depend}
L_{\rm th, \,iso}  
\propto
 {\dot E}_{\rm iso}^{1/3}\,{\bar \Gamma}^{8/3}\, .
\end{equation}

For the non-thermal component, the relation of the peak energy and luminosity to the model parameters 
can be found using the simplest possible description of internal shocks 
that only considers the interaction of two shells of equal mass \citep{barraud_2005}. 
One gets
\begin{equation}
E_{\rm p} 
\propto 
 \dot{E}_{\rm K}^{1/2}\varphi({\cal C})\, \bar{\Gamma}^{-2}\ t_{\rm var}^{-1}\, ,
\end{equation}
\begin{equation}
L_{\rm Nth, \,iso} 
\propto
 f_{\rm IS}\,{\dot E}_{\rm iso}\, ,
\end{equation}
where $t_{\rm var}$ is of the order of one pulse duration, $\varphi({\cal C})$ depends on ${\cal C}$ only, and $f_{\rm IS}$, defined by \eq{eqn_efficiency_is}, is given by 
\begin{equation}
f_{\rm IS}=\epsilon_e\times {1+{\cal C}-2\sqrt{\cal C}\over 1+\cal C}=\epsilon_e\times \epsilon_{\rm diss}({\cal C})\ .
\label{eqn_efficiency_nth}
\end{equation}

It can be seen that the thermal and non-thermal components behave quite differently when the
injected power and average Lorentz factor are changed. This is illustrated in \fig{fig_influence_param} 
where we increase or decrease 
${\dot E}_{\rm iso}$ and ${\bar \Gamma}$ by respective factors of 10 and 2, compared to the reference case shown in
\fig{fig_variable}.

It appears that for a given value of $\epsilon_{\rm th}$ ($\epsilon_{\rm th}=0.03$ in \fig{fig_influence_param}, middle and right panels) 
the thermal component 
becomes more visible when ${\dot E}_{\rm iso}$ is decreased and ${\bar \Gamma}$ increased. This is a direct 
consequence of Eqs.~\ref{eqn_tobs_depend} and \ref{eqn_lth_depend} above, which can be made even more explicit by defining the global thermal
efficiency (assuming $\kappa=0.2 \ \mathrm{cm}^2 \ \mathrm{g}^{-1}$)
\begin{equation}
\label{eqn_efficiency_th}
f_{\rm th}  = {L_{\rm th,iso}\over  {\dot E}_{\rm iso}}
 \simeq 4.9 \times 10^{-3}\, \epsilon_{\rm th}\, (1+\sigma)^{2/3} \, \theta_{-1}^{-2/3}\, l_7^{2/3}\,{\dot E}_{\rm iso, 53}^{-2/3}\,{\bar \Gamma}_2^{8/3}\, ,
\end{equation}
to be compared to the non-thermal efficiency $f_\mathrm{Nth}=f_\mathrm{IS}$ approximated by \eq{eqn_efficiency_nth}. 
While changing  ${\dot E}_{\rm iso}$ and ${\bar \Gamma}$
affects only the thermal efficiency, the opposite is true for the contrast in Lorentz factor ${\cal C}$. Reducing ${\cal C}$
makes internal shocks much less efficient and considerably softens the emitted non-thermal spectrum. In the limit where 
${\cal C} \rightarrow 1$ (and moreover if the Lorentz factor is increasing outwards in the ejecta) there will be no
internal shocks and the emission will only be thermal in the absence of an alternative dissipation process.  

\subsection{Non-thermal emission from magnetic dissipation}
\begin{figure*}
\begin{center}
\begin{tabular}{cc}
\includegraphics[width=0.45\textwidth]{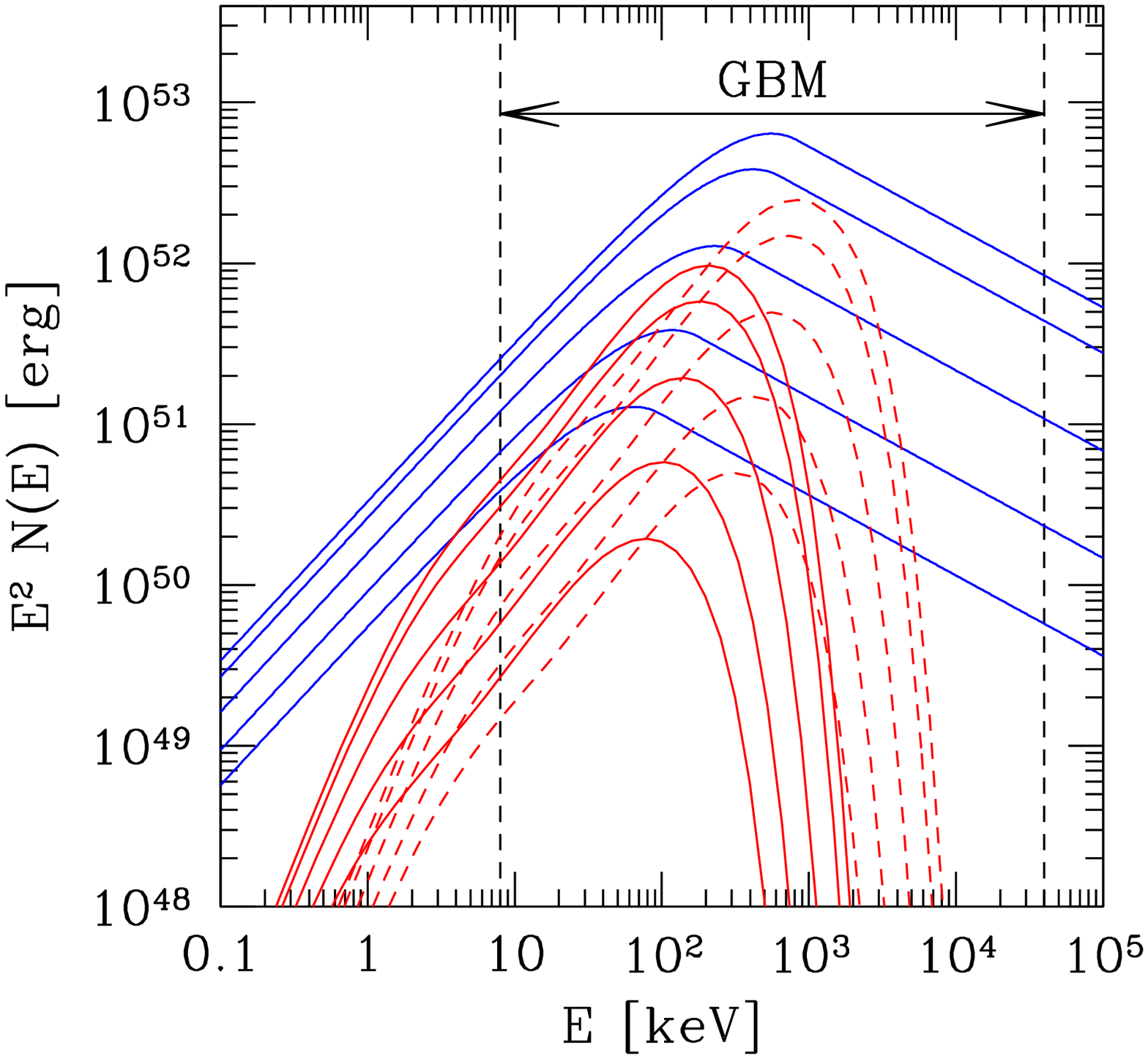} &
\includegraphics[width=0.45\textwidth]{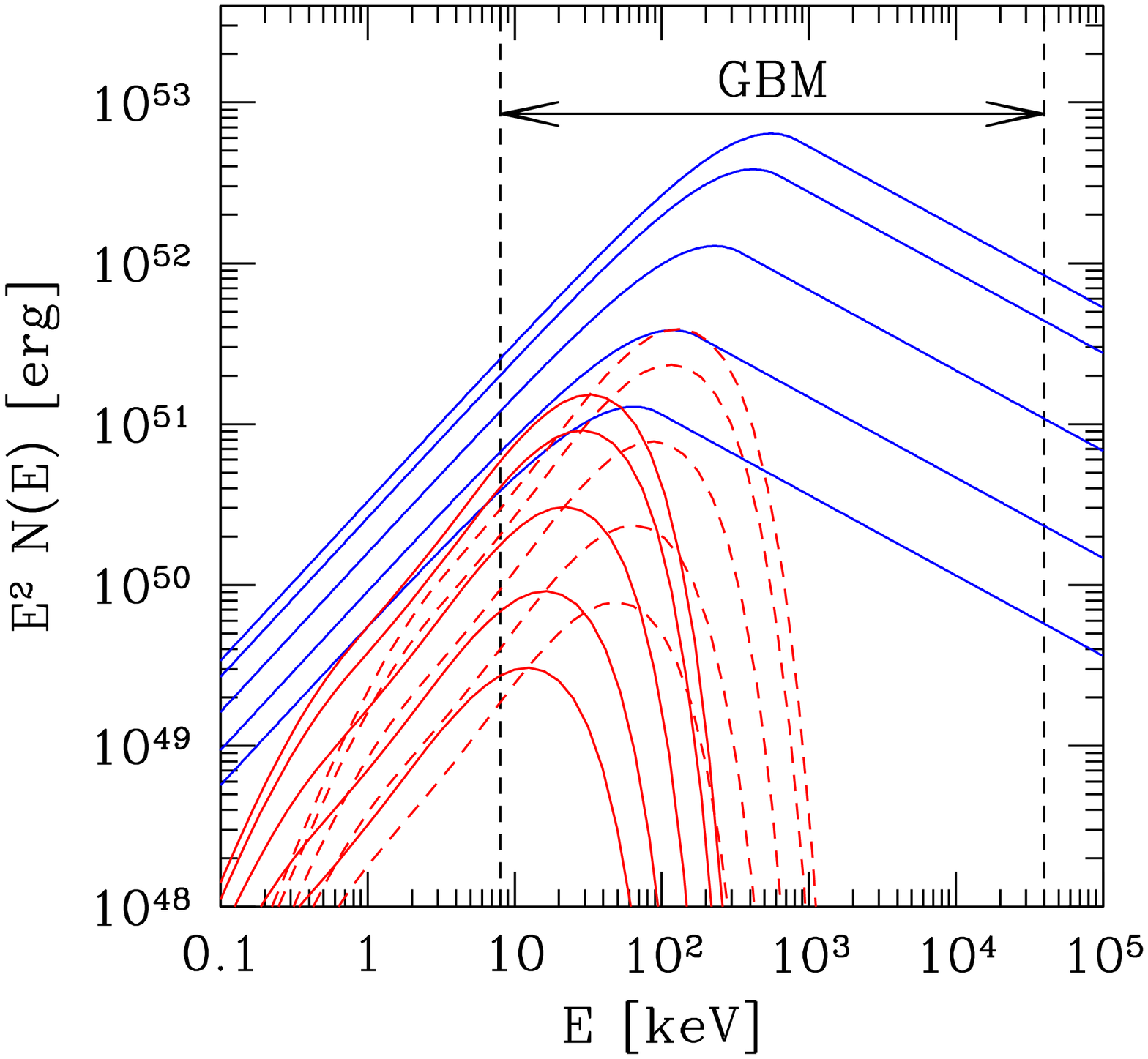}
\end{tabular}
\end{center}

\caption{\textbf{Thermal and non-thermal emission from a variable outflow -- magnetic reconnection framework.}
In each panel we show a sequence of thermal (red) and non-thermal (blue), spectra (the source redshift is $z=1$).
The Lorentz factor distribution adopted for the calculation of the thermal emission is the same as 
in \fig{fig_variable} (left panel) or the same but with $\Gamma$ divided by 2 (right panel).
The thermal emission is computed using the formalism developed in Sect.~\ref{sec_calc_ph}, using five values of  $\epsilon_{\rm th}$: 0.01, 0.03, 0.10, 0.30, and $0.5$. 
The non-thermal spectrum is simply parametrised by a Band function, with the peak energy being given
by the Amati relation (see text) and using five values of the efficiency of the magnetic reconnection, $f_\mathrm{rec}  = f_\mathrm{Nth} (1+\sigma) / \sigma$: 0.01, 0.03, 0.10, 0.30, and $0.5$. 
In all cases, the isotropic magnetic power at the photosphere is fixed to $\sigma \dot{E}_\mathrm{iso} / (1+\sigma) = 10^{53}\, \mathrm{erg\ s^{-1}}$. Finally, the magnetisation $\sigma$ at large distance is either $\sigma=1$ (solid lines) or $10$ (dashed lines).
}
\label{fig_mag}
\end{figure*}

If the non-thermal emission is produced by reconnection in a magnetised outflow, the problem becomes very difficult, 
with no simple way to accurately follow the process
in time and compute a light curve (see e.g. \citealt{spruit_2001,lyutikov_2003,giannios_2008,zhang_2011, mckinney_2012}). As explained in Sect.~2.3 above, we  adopted a very simple assumption to obtain
the non-thermal spectrum: a Band function carrying a fraction $f_{\rm Nth}$ of the injected 
energy with the peak of $E^2 N(E)$ obtained from the Amati relation.
Several examples of the thermal and non-thermal spectra
are represented in the left panel of \fig{fig_mag}, with the thermal component
still being computed with the distribution of Lorentz factor given by \eq{eqn_gp_complex}. 
The results are shown for several values of $\epsilon_\mathrm{th}$ and $\sigma$, and a fixed value of the isotropic magnetic power at the photosphere $\sigma \dot{E}/(1+\sigma)=10^{53}\, \mathrm{erg\ s^{-1}}$.
As expected, the detection of the photospheric component in the spectrum is favored by a high $\epsilon_\mathrm{th}$, a low $f_\mathrm{Nth}$, and a high $\Gamma$. The specific dependency on the magnetisation is discussed in the next section.

\section{Discussion}
\subsection{Relative intensity of the thermal component in GRBs}
Depending on the mechanism responsible for the acceleration of the outflow in 
GRBs, the consequences regarding the photopsheric emission are
very different. In a pure fireball ($\epsilon_{\rm th}=1$) powered by neutrino-antineutrino
annihilation (see e.g. \citealt{popham_1999,zalamea_2011}), the predicted thermal emission is very bright \citep{daigne_2002}. 
To agree with the current data, the spectrum
should then be  Comptonised at high energy (as a result of some dissipative process below the photosphere, e.g. \citealt{rees_2005, peer_2006,giannios_2008,beloborodov_2010,lazzati_2010})
to produce a power law tail and softened at low energy to decrease the spectral index $\alpha$ from
a positive value to a negative one (possibly by the presence of an additional non-thermal component, e.g. \citealt{peer_2006,vurm_2011}). 
An alternative is to suppose that the flow is initially magnetically dominated ($\epsilon_{\rm th}\la 0.1$).
A thermal component is still expected to be released at the photosphere, but it will now be sub-dominant
compared to non-thermal processes such as internal shocks or magnetic reconnection. 

We have explored this second possibility in the present paper, making the following assumptions: 
({\it i}) we supposed that the flow evolves adiabatically from the origin to the photosphere, i.e. we
did not include possible sources of heating below $R_{\rm ph}$; ({\it ii}) if the remaining magnetisation $\sigma$ at the end of acceleration
is weak and does not prevent the formation of internal shocks, we computed their contribution to the emitted radiation as
if $\sigma=0$; ({\it iii}) when $\sigma>1$ we limited ourselves to a very simple parametrised study where we assumed that a fraction
$f_{\rm Nth}$ of the injected power goes into the non-thermal component. 

Regarding the light curve and spectrum of the thermal emission we obtained the following results:
\begin{itemize}
\item Both the photosphere luminosity and temperature depend on the same factor $\Phi$ given by
\eq{eqn_phi}, which directly traces the evolution of the Lorentz factor if the injected power stays
constant. 

\item The duration of the photospheric emission corresponds to the duration $\tau$ of production of the relativistic wind. 
For $t>\tau$ the luminosity drops rapidly on a time scale $\Delta t\sim R_{\rm ph}/2 c \Gamma_{\rm ph}^2\sim$ a few ms
for typical values of the burst parameters.  

\item The global spectrum of the thermal emission is a composite of many elementary contributions
at different temperatures. Before asymptotically reaching a slope $\alpha=+0.4$ at low energy 
it can be much softer below the peak as shown in Fig.~\ref{fig_variable}. 
A time resolved spectrum will resemble more
closely the ``modified Planck function'' adopted for each elementary collision. 
\end{itemize}

If the non-thermal emission comes from internal shocks, its light curve and spectrum  
have been computed using the simplified approach described in Sect.~\ref{sec:nonthermal}.
We have varied several of the model parameters: fraction $\epsilon_{\rm th}$ of thermal energy at the origin of the
flow, isotropic injected power ${\dot E}_{\rm iso}$, and average $\bar \Gamma$
to see under which conditions the thermal component would appear in the observed spectrum,
in the internal shock scenario for $\sigma\la 0.1-1$ (Fig.~\ref{fig_influence_param}) or in the magnetic reconnection scenario for $\sigma \ga 1$ (Fig.~\ref{fig_mag}). 
In the internal shock framework, assuming $\kappa = 0.2 \ \mathrm{cm}^2 \cdot \mathrm{g}^{-1}$, 
the ratio $Q$ of the thermal to non-thermal efficiency is given by
\begin{equation}
\label{eqn_ratio_thnth}
Q={f_{\rm th}\over f_{\rm IS}}=4.9 \times10^{-3}
\frac{\, \epsilon_{\rm th}\, (1+\sigma)^{2/3} \, \theta_{-1}^{-2/3}\, l_7^{2/3}\,{\dot E}_{\rm iso, 53}^{-2/3}\,{\bar \Gamma}_2^{8/3}}
{\epsilon_e\,\epsilon_{\rm diss}({\cal C})}\, .
\end{equation}
As illustrated by this formula and in Fig.~\ref{fig_influence_param},
increasing $\epsilon_{\rm th}$, but also increasing $\bar \Gamma$ or reducing ${\dot E}_{\rm iso}$ or $\cal C$ will
make the thermal component more visible. The magnetisation $\sigma$ has a low impact as $1+\sigma\simeq 1$ in this scenario.
In the examples shown in Fig.~\ref{fig_influence_param},  the model parameters
are taken from the reference case used in \fig{fig_variable} and equal 
$l_{7} = 0.3$,  $\theta_{-1} = 1$, $\epsilon_e=1/3$, and $\sigma = 0.1$.
The effective constrast corresponding to the initial distribution of the Lorentz factor plotted in the upper left panel of \fig{fig_variable} is
${\cal C}\simeq 2.5$.
Then \eq{eqn_ratio_thnth} leads to $Q \simeq  
7.3 \times10^{-2}
\epsilon_{\rm th}\, {\dot E}_{\rm iso, 53}^{-2/3}\, {\bar \Gamma}_2^{8/3}
$, in reasonable agreement with \fig{fig_influence_param}: for instance, the upper left panel corresponds to $\epsilon_\mathrm{th}=0.3$, $\dot{E}_\mathrm{iso, 53}=1$, $\bar{\Gamma}_2\simeq 3$, and $Q\simeq 0.4$, and the bottom right panel corresponds to $\epsilon_\mathrm{th}=0.03$, $\dot{E}_\mathrm{iso, 53}=1$, $\bar{\Gamma}_2 \simeq 1.5$, and $Q\simeq 0.006$. 
Note that in the $E^2 N(E)$ spectrum, the ratio of the maxima of the two components is expected to be slightly higher than $Q$ because the photospheric component has a narrower spectrum than the non-thermal one.

In the case of magnetic reconnection, 
the non-thermal emission is simply parametrised by its global efficiency $f_{\rm Nth}$,
and we have 
\begin{equation}
Q \simeq 0.49 \, \epsilon_{\rm th}\, (1+\sigma)^{2/3} \, \theta_{-1}^{-2/3}\,
l_7^{2/3}\,{\dot E}_{\rm iso, 53}^{-2/3}\,{\bar \Gamma}_2^{8/3}\,f_{\rm Nth,-2}^{-1} \, ,
\label{eq_Qmag}
\end{equation}
in good agreement with \fig{fig_mag} ($f_{\rm Nth,-2}$ being the non-thermal efficiency in \%). Especially, \eq{eq_Qmag} shows that the ratio $Q$ increases with the magnetisation $\sigma$, which seems counter-intuitive. For a fixed value of $\dot{E}$, increasing $\sigma$ reduces $\dot{E}_\mathrm{K}$, and therefore decreases the photospheric radius $R_\mathrm{ph}$ (see \eq{eqn_rph}). Therefore, for a given value of $\epsilon_\mathrm{th}$, the luminosity and temperature of the photosphere increase. However, for a given acceleration mechanism, one would expect an increase of $\sigma$ to be associated with a decrease of $\epsilon_\mathrm{th}$, which may affect the dependency of the ratio $Q$ on the magnetisation $\sigma$. For instance, in the case of a passive magnetic field, including $\sigma=\sigma_\mathrm{passive}$ in \eq{eq_Qmag} leads to 
$
Q \simeq 0.49 \,  (1+\sigma)^{-1/3} \, \theta_{-1}^{-2/3}\,
l_7^{2/3}\,{\dot E}_{\rm iso, 53}^{-2/3}\,{\bar \Gamma}_2^{8/3}\,f_{\rm Nth,-2}^{-1}
$, i.e. a decreasing ratio for an increasing magnetisation. 
Since increasing the final magnetisation $\sigma$ tends to decrease the photospheric radius, 
it should also be noted that the acceleration of the flow may well be incomplete at the photosphere in 
high $\sigma$ scenarios. As discussed in appendix~\ref{sec_appendix}, this will also reduce the photospheric emission.\\

\subsection{Comparison to observations}

We now check how these results compare to the various observational indications of the presence of a thermal
component in GRB spectra.\\

One first indirect indication comes from the very hard spectral
slopes $\alpha >0$ that are sometimes observed during burst evolution (e.g. \citealt{ghirlanda_2003, bosnjak_2006, abdo_2009}). Indeed, our results allow
the thermal over non-thermal ratio (\eq{eqn_ratio_thnth}) to vary with time. A locally large $\bar \Gamma$ will boost the
thermal component while a low contrast $\cal C$ will reduce the non-thermal component (under the condition that the
non-thermal emission comes from internal shocks). This can explain an erratic behaviour of the $\alpha$ slope,
but observations often show a regular shift of $\alpha$ from positive to negative values during a single pulse.
As explained in Sect.~3.1 a thermal start and a non-thermal ending are predicted by our models, but the fraction
of time during which the emission is thermal is generally smaller than observed in the few bursts where the measured $\alpha$ slope is positive. It remains possible to adopt a 
distribution of the Lorentz factor that would extend the duration of the thermal emission but smoothly connecting the thermal and
non-thermal components may not be easy. \\

\begin{figure*}
\centerline{\includegraphics[width=0.71\textwidth]{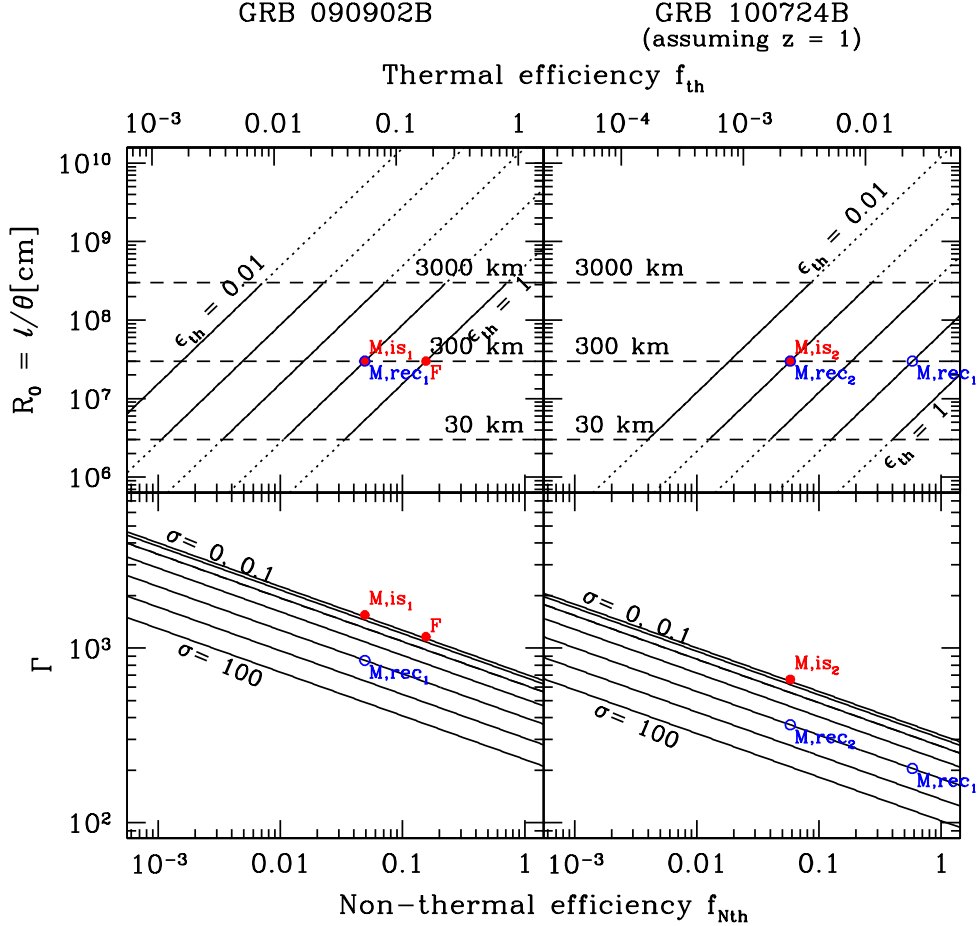}}

\caption{\textbf{Constraints on the thermal and non-thermal emission in GRB 090902B and GRB 100724B.} \textit{Top:} for a given thermal fraction $\epsilon_\mathrm{th}=10^{-2}$, $10^{-1.5}$, $10^{-1}$, $10^{-0.5}$, and $1$, the radius $R_0=\ell/\theta$ at the base of the flow is plotted as a function of the non-thermal efficiency $f_\mathrm{Nth}$. The corresponding thermal efficiency $f_\mathrm{th}$ is also shown (top $x$-axis). \textit{Bottom:} for a given magnetisation $\sigma=10^{-1}$, $10^{-0.5}$, $1$, $10^{0.5}$, $10^1$, $10^{1.5}$, and $10^2$ at the end of acceleration phase, the Lorentz factor of the flow is plotted as a function of $f_\mathrm{Nth}$ (the unmagnetised case $\sigma=0$ cannot be distinguished from the case $\sigma=0.1$). 
Sets of parameters representative of the different classes of scenarios discussed in the paper are indicated: F ($\epsilon_\mathrm{th}=1$, $\sigma=0$) (standard fireball), M,is$_\mathrm{1}$ ($\log{\epsilon_\mathrm{th}}=-0.5$, $\sigma=0$,) and M,is$_\mathrm{2}$ ($\log{\epsilon_\mathrm{th}}=-1.5$, $\sigma=0$) (efficient magnetic acceleration: magnetisation is low above the photosphere and the dominant non-thermal mechanism is internal shocks), M,rec$_\mathrm{1}$ ($\log{\epsilon_\mathrm{th}}=-0.5$, $\sigma=10$), and M,rec$_\mathrm{2}$ ($\log{\epsilon_\mathrm{th}}=-1.5$, $\sigma=10$) (magnetised flow at large distance, the dominant non-thermal mechanism is magnetic reconnection). The initial radius is fixed to $R_0=300\, \mathrm{km}$, a typical value  for long GRBs. 
The observational data (thermal flux, temperature, ratio of the thermal over the total flux) used for the calculation (see text) are taken from \citet{abdo_2009,peer_2012} for GRB 090902B (left column), and from \citet{guiriec_2011} for GRB 100724B (right column).
}
\label{fig:modelobs}
\end{figure*}

A second indication comes from the recent detection of possible thermal components in a few \textit{Fermi} GRBs.
GRB 090902B is a very peculiar case of a burst showing a spectrum 
best fitted by a Band function with a 
hard
low-energy photon index, or a multi-colour black body, together with an additional sub-dominant power law
 \citep{abdo_2009,ryde_2010}. A first possibility is to interpret this burst
 as photospheric emission with additional dissipative processes affecting the spectrum by Comptonisation and possibly other non-thermal processes. Conversely, using the formalism presented here, we can also derive constraints on the burst parameters for alternative scenarios where the Band component is interpreted as the thermal emission produced at the photosphere without sub-photospheric dissipation, and the power law as non-thermal emission produced above the photosphere in the optically thin regime.
From the analysis made by \citet{abdo_2009} and \citet{peer_2012}, the implied thermal luminosity and temperature are large, 
$L_{\rm th}\simeq 4.6\times 10^{53} 
\, \mathrm{erg.s^{-1}}$
and 
$T_\mathrm{obs}\simeq 168
\,\mathrm{keV}$.
The thermal luminosity represents  $53 \%$
of the total luminosity.
Then, for a given $\epsilon_\mathrm{th}$, 
$\sigma$,
and $f_\mathrm{Nth}$, one can deduce from Eqs.~\ref{eqn_tobs}, \ref{eqn_lth}, and \ref{eqn_rph} the isotropic power $\dot{E}_{\rm iso}$, the size $R_0=\ell/\theta$ of the region at the base of the outflow (see \fig{fig:modelobs}, top-left panel), and the Lorentz factor $\Gamma$ (see \fig{fig:modelobs}, bottom-left panel). This is a similar approach to the one proposed by \citet{peer_2007}, but within the more general framework defined in this paper, which allows us to consider several scenarios for the acceleration of the outflow.
As GRB 090902B is very bright, this  leads to huge values of $\dot{E}_{\rm iso}$, the minimum being obtained for $\epsilon_\mathrm{th}\simeq 1$, which would make GRB 090902B a peculiar  burst associated to a situation close to a pure fireball. For instance, for  $\epsilon_{\rm th} = 1$, $l_{7} = 0.3$, $\theta_{-1} = 1$ (i.e. $R_0=300\, \mathrm{km}$), and $\sigma \ll 1$ we get an isotropic power $\dot{E}_{\rm iso}\sim 
2.6\times 10^{54}\, \mathrm{erg.s^{-1}}$ and a Lorentz factor $\Gamma\sim 1160$. The efficiency of the photospheric emission in this case is $f_{\rm th} \simeq 0.18$ and the efficiency of the non-thermal emission above the photosphere is $f_\mathrm{Nth}\simeq 0.16$, marginally compatible with internal shocks. This case is labelled as F in \fig{fig:modelobs} (left column) and is in good agreement with the analysis made by \citet{peer_2012}.

As most GRBs do not show such a bright thermal component, they usually require much lower values of $\epsilon_\mathrm{th}$, as illustrated below with GRB 100724B. Then, it is worth  studying the possibility to model GRB 090902B with $\epsilon_\mathrm{th}<1$, which would correspond to an initially magnetised outflow. Reducing $\epsilon_\mathrm{th}$ in this burst
has several consequences. As illustrated in \fig{fig:modelobs} (left column), it implies a general decrease of the efficiency, except if the initial size $R_0=\ell/\theta$ is very large ($R_0 \gg 3000$ km), which is not expected for most models of the central engine of long GRBs, because of their short timescale ($\sim 1-10\, \mathrm{ms}$) variability.
For instance, for $\epsilon_\mathrm{th}=10^{-0.5}$ (case labelled as M,is$_\mathrm{1}$, where we keep $R_0=300$ km and $\sigma\ll1$), the thermal and non-thermal efficiencies are reduced to $f_\mathrm{th}=0.06$ and $f_\mathrm{Nth}=0.05$, leading to an isotropic equivalent power $\dot{E}_\mathrm{iso}=9.4\times 10^{54}\, \mathrm{erg.s^{-1}}$. It also leads to an increase of the Lorentz factor, which was already quite high for the pure fireball scenario (see \fig{fig:modelobs}, bottom-left panel). Therefore, $\epsilon_\mathrm{th}\simeq 0.3-0.5$ seems a reasonable lower limit for GRB 090902B. Note that a decrease of $\epsilon_\mathrm{th}$ compared to the standard fireball leads in this case to a reduced non-thermal efficiency $f_\mathrm{th}$ in good agreement with the expected value for internal shocks. For higher values of $\epsilon_\mathrm{th}$, magnetic reconnection, which is supposed to have a higher efficiency, is a better candidate, as illustrated in  \fig{fig:modelobs}  (left column). In addition, for a fixed value of $\epsilon_\mathrm{th}$, increasing the magnetisation $\sigma$ at the end of acceleration  always reduces the constraint on the Lorentz factor (bottom left panel). A more detailed modelling of GRB 090902B would be necessary to distinguish between these different possibilities. \\

Conversely, the results of \citet{guiriec_2011} indicating the presence of a sub-dominant
thermal component in GRB 100724B, representing about 4\% of the total flux, point towards low values of $\epsilon_{\rm th}$ 
and, therefore,
a magnetic acceleration.  
As shown in \fig{fig:modelobs}, it is difficult to interpret GRB 100724B  within the standard fireball scenario (thermal acceleration) that would imply either $f_\mathrm{Nth} > 1$ or $R_0 < 120\, \mathrm{km}$ (and even $R_0<40\, \mathrm{km}$ if $f_\mathrm{Nth}<0.5$ is required). This is obtained assuming a redshift $z=1$, but we checked that our conclusions are unchanged for larger redshifts. Then a low thermal fraction $\epsilon_\mathrm{th} \la 0.01-0.1$ is required in GRB 100724B, whose spectral properties are much more representative of the bulk GRB population than in the unusual case of GRB 090902B.
A similar conclusion was obtained by \citet{zhang_2009} in the case of the very energetic burst GRB 080916C where no bright thermal component was detected. Assuming a passive magnetic field below the photosphere, they obtain the constraint $\sigma_\mathrm{passive}\ga 15-20$, which leads to the more general 
condition $\epsilon_\mathrm{th}\la 0.05$, from Eq.~\ref{eq:signa_passive}.

If magnetic acceleration is common in GRBs, several scenarios can be discussed.
In scenarios where magnetic acceleration is efficient, implying a low magnetisation at large radius and a dominant role of internal shocks, the thermal fraction should not be much larger than a few percents to avoid an unrealistic efficiency $f_\mathrm{Nth}$. 
For instance in the case of GRB 100724B, for $z=1$, $\epsilon_\mathrm{th}=10^{-1.5}$, $\ell_7=0.3$, $\theta_{-1}=1$ (i.e. $R_0= 300\, \mathrm{km}$), and $\sigma\ll 1$ (case labelled as M,is$_2$ in \fig{fig:modelobs} right column), we get a non-thermal efficiency $f_\mathrm{Nth}\simeq 0.06$ and a thermal efficiency $f_\mathrm{th}\simeq 2\times 10^{-3}$. The isotropic kinetic power and the Lorentz factor in this case are $\dot{E}_\mathrm{iso}\simeq 5.6\times 10^{53}\, \mathrm{erg.s^{-1}}$ and $\Gamma\simeq 660$. Alternative scenarios -- where the flow is still magnetised at large radius and magnetic reconnection is the dominant mechanism to produce non-thermal emission -- are less constrained, because of the uncertainties in the underlying physics. As illustrated in \fig{fig:modelobs}, for a fixed $\epsilon_\mathrm{th}$, increasing $\sigma$ tends to reduce $\Gamma$, which is already in the typical range of a few hundreds  for $\sigma=0$. High values of the non-thermal efficiency $f_\mathrm{Nth}\ga 0.1-0.5$ (as usually expected for magnetic reconnection, see e.g. \citealt{zhang_2011,mckinney_2012}) also require high values of the thermal fraction $\epsilon_\mathrm{th}\ga 0.1-0.3$, which does not seem natural in such scenarios of highly magnetised outflows.   

As illustrated in Sect.~\ref{sec:results}, spectra
with non-thermal and thermal components resembling those 
found by \citet{guiriec_2011} are easily obtained with our model, either for 
a photospheric + internal shocks scenario in a case of efficient magnetic acceleration, or 
for a photospheric + reconnection scenario if the magnetisation at large distance is still large.
A potential issue for the internal shock scenario is the moderate variation of the temperature (within a factor of 2) found in the time-resolved ana\-lysis. 
To be efficient, internal shocks require 
large fluctuations of the Lorentz factors that are even amplified in the observed temperature ($T_{\rm obs}\propto \Gamma^{8/3}$, see \fig{fig_variable}).
This may suggest that the non-thermal emission in GRB 100724B comes from magnetic reconnection. This is unfortunately difficult to test in absence of theoretical predictions for the spectral evolution in this case.
It should however be
noted that when the temperature drops, the luminosity also drops so that, in practice, 
the temperature can be determined only 
when it is high enough. Depending on the time scale for the Lorentz factor fluctuations and 
the temporal resolution of the analysis, this may artificially reduce the amplitude of the 
measured variations of temperature. This is illustrated in \fig{fig_variable} where dotted and dashed
lines show the temperature (and peak energy of the non-thermal spectrum) averaged over
intervals of 2 and 4 s, respectively.              
It remains to be tested if this smoothing effect can account for GRB 100724B evolution in the photospheric + internal shocks scenario.

\section{Conclusion}

We have explored in detail GRB scenarios with two episodes of emission: thermal emission from the photosphere without sub-photospheric dissipation, and non-thermal emission from internal dissipation above the photosphere.
Our results can be used to interpret the data and 
obtain constraints on the burst parameters or acceleration mechanism. 
But one faces the difficulty arising from the diversity of the proposed evidence 
for the presence of a thermal component in GRB spectra. In some cases this thermal component 
represents a major contribution to the global spectrum (with additional non-thermal contributions)
while in others it is always sub-dominant, most of the emission having a non-thermal origin. These different
situations seem to imply quite different magnetic over thermal energy ratios at the origin of the flow.
However the lack of bright thermal components in most GRBs clearly points towards magnetic acceleration, with $\epsilon_{\rm th} \la 0.01$ in most cases, and $\epsilon_{\rm th} \simeq 0.01-0.1$ in less frequent cases such as GRB 100724B.
GRB 090902B with $\epsilon_{\rm th} \simeq 0.3-1$ remains an exception.

More generally, one may wonder what would be the best conditions for the thermal emission to show up.
Apart from the obvious requirement that $\epsilon_{\rm th}$ should be as large as possible, \eq{eqn_ratio_thnth} may
suggest looking for events with a low $\dot{E}_{\rm iso}$ and/or a large average Lorentz factor. This, however, 
supposes that these two quantities are independent. Having $\Gamma\propto \dot{E}_{\rm iso}^{q}$ and $q>1/4$ 
would favor both a large $\dot{E}_{\rm iso}$ and $\Gamma$ while the opposite is true for $q<1/4$. 
Finally, if the non-thermal emission comes from internal shocks, a pure thermal spectrum can be
possible even if the distribution of the Lorentz factor has a low contrast or if $\Gamma$ is increasing 
outwards. 

The observation of a burst with an unambiguous photospheric signature in its spectrum would
greatly help to clarify several issues in GRB physics: {\it (i)} estimating the value of $\epsilon_{\rm th}$ would
provide insight on the acceleration mechanism of the flow; {\it (ii)} obtaining the temperature and thermal luminosity
evolution would constrain the distribution of Lorentz factor and injected power; and {\it (iii)} measuring the level 
of temperature fluctuations with a high temporal resolution 
would help to discriminate between internal shocks and magnetic reconnection 
for the non-thermal emission. 
Isolating the photospheric component in the available data is, however, not an easy task:
it is generally one among other spectral components and possibly sub-dominant, and does not have a simple blackbody spectrum.

\appendix
\section{Incomplete acceleration at the photosphere}
\label{sec_appendix}
In the case where the flow is still accelerating at the photosphere, the expressions for
the photospheric radius, observed tempe\-rature, and thermal luminosity will depend both
on the Lorentz factor at the photosphere and on its value at the
end of acceleration $\Gamma_{\infty}$.
To obtain the new expressions for $R_{\rm ph}$, $T_{\rm obs}$, and $L_{\rm th}$  we write  
that the optical depth seen by a photon produced at $R_{\rm ph}$ and leaving
the flow at $R_{\rm out}$ is equal to unity
\begin{equation}
\tau=\int_{R_{\rm ph}}^{R_{\rm out}} {\kappa\,{\dot M}\over 8\pi\,c\,\Gamma^2\,R^2}\,dr=1\ .
\end{equation}
For simplicity we suppose in this appendix that the flow is stationary, i.e. that ${\dot M}$ is constant and 
that $\Gamma_{\infty}$ is identical for all the shells.
Then, adopting a simple parametrisation for the Lorentz factor 
\begin{equation}
\Gamma = \left\lbrace\begin{array}{cl}
& \Gamma_{\rm ph}\left({R\over R_{\rm ph}}\right)^{\alpha}\ \ \ {\rm for}\ \ R_{\rm ph}<R<R_\mathrm{sat}\ \ \ \ (\alpha>0)\\
& \Gamma_{\infty}\ \ \ {\rm for}\ \ R>R_\mathrm{sat}\\
\end{array}\right.
\end{equation}
and with ${\dot M}={\dot E}_{\rm K}/\Gamma_{\infty}\,c^2$ we get (neglecting terms of the order of
$(R_{\rm ph}/R_\mathrm{sat})^{2\alpha+1}$)
\begin{equation}
\tau\approx {\kappa\,{\dot E}_{\rm K}\over 8\pi\,(2\alpha+1)\,c^3\,\Gamma_{\infty}\Gamma_{\rm ph}^2\,R_{\rm ph}}\ .
\end{equation} 
This finally leads to
\begin{equation}
{R_{\rm ph}\over R_{\rm ph,c}}={1\over 2\alpha+1}\left({\Gamma_{\infty}\over \Gamma_{\rm ph}}\right)^2
\end{equation}
and
\begin{equation}
{T_{\rm obs}\over T_{\rm obs,c}}={L_{\rm th}\over L_{\rm th,c}}=
({2\alpha+1})^{2/3}\left({\Gamma_{\rm ph}\over \Gamma_{\infty}}\right)^2=
({2\alpha+1})^{2/3}\left({R_{\rm ph}\over R_\mathrm{sat}}\right)^{2\alpha}\!\! ,
\end{equation}
where the index $c$ refers to the case where the acceleration is essentially complete at the 
photosphere. It can be seen that an incomplete acceleration can substantially reduce the thermal
contribution. 
\begin{acknowledgements} 
The authors thank the referee for constructive comments helping us to clarify the formulation of the paper.
This work is partially supported by a grant from the French Space Agency (CNES). 
R.H.'s PhD work is funded by a Fondation CFM-JP Aguilar grant.
\end{acknowledgements}\vspace*{-3ex}

\bibliographystyle{aa} 
\bibliography{t}

\begin{thebibliography}{78}
\expandafter\ifx\csname natexlab\endcsname\relax\def\natexlab#1{#1}\fi

\bibitem[{{Abdo} {et~al.}(2009){Abdo}, {Ackermann}, {Ajello}, {Asano},
  {Atwood}, {Axelsson}, {Baldini}, {Ballet}, {Barbiellini}, {Baring},
  {Bastieri}, {Bechtol}, {Bellazzini}, {Berenji}, {Bhat}, {Bissaldi},
  {Blandford}, {Bloom}, {Bonamente}, {Borgland}, {Bouvier}, {Bregeon}, {Brez},
  {Briggs}, {Brigida}, {Bruel}, {Burgess}, {Burrows}, {Buson}, {Caliandro},
  {Cameron}, {Caraveo}, {Casandjian}, {Cecchi}, {{\c C}elik}, {Chekhtman},
  {Cheung}, {Chiang}, {Ciprini}, {Claus}, {Cohen-Tanugi}, {Cominsky},
  {Connaughton}, {Conrad}, {Cutini}, {d'Elia}, {Dermer}, {de Angelis}, {de
  Palma}, {Digel}, {Dingus}, {Silva}, {Drell}, {Dubois}, {Dumora}, {Farnier},
  {Favuzzi}, {Fegan}, {Finke}, {Fishman}, {Focke}, {Fortin}, {Frailis},
  {Fukazawa}, {Funk}, {Fusco}, {Gargano}, {Gehrels}, {Germani}, {Giavitto},
  {Giebels}, {Giglietto}, {Giordano}, {Glanzman}, {Godfrey}, {Goldstein},
  {Granot}, {Greiner}, {Grenier}, {Grove}, {Guillemot}, {Guiriec}, {Hanabata},
  {Harding}, {Hayashida}, {Hays}, {Horan}, {Hughes}, {Jackson},
  {J{\'o}hannesson}, {Johnson}, {Johnson}, {Johnson}, {Kamae}, {Katagiri},
  {Kataoka}, {Kawai}, {Kerr}, {Kippen}, {Kn{\"o}dlseder}, {Kocevski}, {Komin},
  {Kouveliotou}, {Kuss}, {Lande}, {Latronico}, {Lemoine-Goumard}, {Longo},
  {Loparco}, {Lott}, {Lovellette}, {Lubrano}, {Madejski}, {Makeev},
  {Mazziotta}, {McBreen}, {McEnery}, {McGlynn}, {Meegan}, {M{\'e}sz{\'a}ros},
  {Meurer}, {Michelson}, {Mitthumsiri}, {Mizuno}, {Moiseev}, {Monte},
  {Monzani}, {Moretti}, {Morselli}, {Moskalenko}, {Murgia}, {Nakamori},
  {Nolan}, {Norris}, {Nuss}, {Ohno}, {Ohsugi}, {Omodei}, {Orlando}, {Ormes},
  {Paciesas}, {Paneque}, {Panetta}, {Pelassa}, {Pepe}, {Pesce-Rollins},
  {Petrosian}, {Piron}, {Porter}, {Preece}, {Rain{\`o}}, {Rando}, {Rau},
  {Razzano}, {Razzaque}, {Reimer}, {Reimer}, {Reposeur}, {Ritz}, {Rochester},
  {Rodriguez}, {Roming}, {Roth}, {Ryde}, {Sadrozinski}, {Sanchez}, {Sander},
  {Saz Parkinson}, {Scargle}, {Schalk}, {Sgr{\`o}}, {Siskind}, {Smith},
  {Spinelli}, {Stamatikos}, {Stecker}, {Stratta}, {Strickman}, {Suson},
  {Swenson}, {Tajima}, {Takahashi}, {Tanaka}, {Thayer}, {Thayer}, {Thompson},
  {Tibaldo}, {Torres}, {Tosti}, {Tramacere}, {Uchiyama}, {Uehara}, {Usher},
  {van der Horst}, {Vasileiou}, {Vilchez}, {Vitale}, {von Kienlin}, {Waite},
  {Wang}, {Wilson-Hodge}, {Winer}, {Wood}, {Yamazaki}, {Ylinen}, \&
  {Ziegler}}]{abdo_2009}
{Abdo}, A.~A., {Ackermann}, M., {Ajello}, M., {et~al.} 2009, \apjl, 706, L138

\bibitem[{{Amati} {et~al.}(2002){Amati}, {Frontera}, {Tavani}, {in't Zand},
  {Antonelli}, {Costa}, {Feroci}, {Guidorzi}, {Heise}, {Masetti}, {Montanari},
  {Nicastro}, {Palazzi}, {Pian}, {Piro}, \& {Soffitta}}]{amati_2002}
{Amati}, L., {Frontera}, F., {Tavani}, M., {et~al.} 2002, \aap, 390, 81

\bibitem[{{Band} {et~al.}(1993){Band}, {Matteson}, {Ford}, {Schaefer},
  {Palmer}, {Teegarden}, {Cline}, {Briggs}, {Paciesas}, {Pendleton}, {Fishman},
  {Kouveliotou}, {Meegan}, {Wilson}, \& {Lestrade}}]{band_1993}
{Band}, D., {Matteson}, J., {Ford}, L., {et~al.} 1993, \apj, 413, 281

\bibitem[{{Band} \& {Preece}(2005)}]{band_2005}
{Band}, D.~L. \& {Preece}, R.~D. 2005, \apj, 627, 319

\bibitem[{{Barraud} {et~al.}(2005){Barraud}, {Daigne}, {Mochkovitch}, \&
  {Atteia}}]{barraud_2005}
{Barraud}, C., {Daigne}, F., {Mochkovitch}, R., \& {Atteia}, J.~L. 2005, \aap,
  440, 809

\bibitem[{{Begelman} \& {Li}(1994)}]{begelman_1994}
{Begelman}, M.~C. \& {Li}, Z.-Y. 1994, \apj, 426, 269

\bibitem[{{Beloborodov}(2010)}]{beloborodov_2010}
{Beloborodov}, A.~M. 2010, \mnras, 407, 1033

\bibitem[{{Beloborodov}(2011)}]{beloborodov_2011}
{Beloborodov}, A.~M. 2011, \apj, 737, 68

\bibitem[{{Bosnjak} {et~al.}(2006){Bosnjak}, {Celotti}, \&
  {Ghirlanda}}]{bosnjak_2006}
{Bosnjak}, Z., {Celotti}, A., \& {Ghirlanda}, G. 2006, \mnras, 370, L33

\bibitem[{{Bo{\v s}njak} {et~al.}(2009){Bo{\v s}njak}, {Daigne}, \&
  {Dubus}}]{bosnjak_2009}
{Bo{\v s}njak}, {\v Z}., {Daigne}, F., \& {Dubus}, G. 2009, \aap, 498, 677

\bibitem[{{Brainerd} \& {Lamb}(1987)}]{brainerd_1987}
{Brainerd}, J.~J. \& {Lamb}, D.~Q. 1987, \apj, 313, 231

\bibitem[{{Cline} \& {Desai}(1975)}]{cline_1975}
{Cline}, T.~L. \& {Desai}, U.~D. 1975, \apjl, 196, L43

\bibitem[{{Cline} {et~al.}(1973){Cline}, {Desai}, {Klebesadel}, \&
  {Strong}}]{cline_1973}
{Cline}, T.~L., {Desai}, U.~D., {Klebesadel}, R.~W., \& {Strong}, I.~B. 1973,
  \apjl, 185, L1

\bibitem[{{Collazzi} {et~al.}(2012){Collazzi}, {Schaefer}, {Goldstein}, \&
  {Preece}}]{collazzi_2012}
{Collazzi}, A.~C., {Schaefer}, B.~E., {Goldstein}, A., \& {Preece}, R.~D. 2012,
  \apj, 747, 39

\bibitem[{{Daigne} {et~al.}(2011){Daigne}, {Bo{\v s}njak}, \&
  {Dubus}}]{daigne_2011}
{Daigne}, F., {Bo{\v s}njak}, {\v Z}., \& {Dubus}, G. 2011, \aap, 526, A110

\bibitem[{{Daigne} \& {Drenkhahn}(2002)}]{daigne_2002b}
{Daigne}, F. \& {Drenkhahn}, G. 2002, \aap, 381, 1066

\bibitem[{{Daigne} \& {Mochkovitch}(1998)}]{daigne_1998}
{Daigne}, F. \& {Mochkovitch}, R. 1998, \mnras, 296, 275

\bibitem[{{Daigne} \& {Mochkovitch}(2002)}]{daigne_2002}
{Daigne}, F. \& {Mochkovitch}, R. 2002, \mnras, 336, 1271

\bibitem[{{Derishev} {et~al.}(2001){Derishev}, {Kocharovsky}, \&
  {Kocharovsky}}]{derishev_2001}
{Derishev}, E.~V., {Kocharovsky}, V.~V., \& {Kocharovsky}, V.~V. 2001, \aap,
  372, 1071

\bibitem[{{Ghirlanda} {et~al.}(2003){Ghirlanda}, {Celotti}, \&
  {Ghisellini}}]{ghirlanda_2003}
{Ghirlanda}, G., {Celotti}, A., \& {Ghisellini}, G. 2003, \aap, 406, 879

\bibitem[{{Ghirlanda} {et~al.}(2012){Ghirlanda}, {Ghisellini}, {Nava},
  {Salvaterra}, {Tagliaferri}, {Campana}, {Covino}, {D'Avanzo}, {Fugazza},
  {Melandri}, \& {Vergani}}]{ghirlanda_2012}
{Ghirlanda}, G., {Ghisellini}, G., {Nava}, L., {et~al.} 2012, \mnras, 422, 2553

\bibitem[{{Ghisellini} {et~al.}(2000){Ghisellini}, {Celotti}, \&
  {Lazzati}}]{ghisellini_2000}
{Ghisellini}, G., {Celotti}, A., \& {Lazzati}, D. 2000, \mnras, 313, L1

\bibitem[{{Giannios}(2008)}]{giannios_2008}
{Giannios}, D. 2008, \aap, 480, 305

\bibitem[{{Giannios}(2012)}]{giannios_2012}
{Giannios}, D. 2012, \mnras, 422, 3092

\bibitem[{{Giannios} \& {Spruit}(2007)}]{giannios_2007}
{Giannios}, D. \& {Spruit}, H.~C. 2007, \aap, 469, 1

\bibitem[{{Gilman} {et~al.}(1980){Gilman}, {Metzger}, {Parker}, {Evans}, \&
  {Trombka}}]{gilman_1980}
{Gilman}, D., {Metzger}, A.~E., {Parker}, R.~H., {Evans}, L.~G., \& {Trombka},
  J.~I. 1980, \apj, 236, 951

\bibitem[{{Goldstein} {et~al.}(2012){Goldstein}, {Burgess}, {Preece}, {Briggs},
  {Guiriec}, {van der Horst}, {Connaughton}, {Wilson-Hodge}, {Paciesas},
  {Meegan}, {von Kienlin}, {Bhat}, {Bissaldi}, {Chaplin}, {Diehl}, {Fishman},
  {Fitzpatrick}, {Foley}, {Gibby}, {Giles}, {Greiner}, {Gruber}, {Kippen},
  {Kouveliotou}, {McBreen}, {McGlynn}, {Rau}, \& {Tierney}}]{goldstein_2012}
{Goldstein}, A., {Burgess}, J.~M., {Preece}, R.~D., {et~al.} 2012, \apjs, 199,
  19

\bibitem[{{Goodman}(1986)}]{goodman_1986}
{Goodman}, J. 1986, \apjl, 308, L47

\bibitem[{{Granot} {et~al.}(2011){Granot}, {Komissarov}, \&
  {Spitkovsky}}]{granot_2011}
{Granot}, J., {Komissarov}, S.~S., \& {Spitkovsky}, A. 2011, \mnras, 411, 1323

\bibitem[{{Guiriec} {et~al.}(2011){Guiriec}, {Connaughton}, {Briggs},
  {Burgess}, {Ryde}, {Daigne}, {M{\'e}sz{\'a}ros}, {Goldstein}, {McEnery},
  {Omodei}, {Bhat}, {Bissaldi}, {Camero-Arranz}, {Chaplin}, {Diehl}, {Fishman},
  {Foley}, {Gibby}, {Giles}, {Greiner}, {Gruber}, {von Kienlin}, {Kippen},
  {Kouveliotou}, {McBreen}, {Meegan}, {Paciesas}, {Preece}, {Rau}, {Tierney},
  {van der Horst}, \& {Wilson-Hodge}}]{guiriec_2011}
{Guiriec}, S., {Connaughton}, V., {Briggs}, M.~S., {et~al.} 2011, \apjl, 727,
  L33

\bibitem[{{Hasco{\"e}t} {et~al.}(2012){Hasco{\"e}t}, {Daigne}, \&
  {Mochkovitch}}]{hascoet_2012}
{Hasco{\"e}t}, R., {Daigne}, F., \& {Mochkovitch}, R. 2012, \aap, 542, L29

\bibitem[{{Kaneko} {et~al.}(2006){Kaneko}, {Preece}, {Briggs}, {Paciesas},
  {Meegan}, \& {Band}}]{kaneko_2006}
{Kaneko}, Y., {Preece}, R.~D., {Briggs}, M.~S., {et~al.} 2006, \apjs, 166, 298

\bibitem[{{Kobayashi} {et~al.}(1997){Kobayashi}, {Piran}, \&
  {Sari}}]{kobayashi_1997}
{Kobayashi}, S., {Piran}, T., \& {Sari}, R. 1997, \apj, 490, 92

\bibitem[{{Kocevski}(2012)}]{kocevski_2012}
{Kocevski}, D. 2012, \apj, 747, 146

\bibitem[{{Komissarov} {et~al.}(2010){Komissarov}, {Vlahakis}, \&
  {K{\"o}nigl}}]{komissarov_2010}
{Komissarov}, S.~S., {Vlahakis}, N., \& {K{\"o}nigl}, A. 2010, \mnras, 407, 17

\bibitem[{{Komissarov} {et~al.}(2009){Komissarov}, {Vlahakis}, {K{\"o}nigl}, \&
  {Barkov}}]{komissarov_2009}
{Komissarov}, S.~S., {Vlahakis}, N., {K{\"o}nigl}, A., \& {Barkov}, M.~V. 2009,
  \mnras, 394, 1182

\bibitem[{{Lazzati} \& {Begelman}(2010)}]{lazzati_2010}
{Lazzati}, D. \& {Begelman}, M.~C. 2010, \apj, 725, 1137

\bibitem[{{Lyutikov} \& {Blandford}(2003)}]{lyutikov_2003}
{Lyutikov}, M. \& {Blandford}, R. 2003, ArXiv Astrophysics e-prints

\bibitem[{{Mazets} {et~al.}(1981){Mazets}, {Golenetskii}, {Ilinskii}, {Panov},
  {Aptekar}, {Gurian}, {Proskura}, {Sokolov}, {Sokolova}, \&
  {Kharitonova}}]{mazets_1981}
{Mazets}, E.~P., {Golenetskii}, S.~V., {Ilinskii}, V.~N., {et~al.} 1981, \apss,
  80, 3

\bibitem[{{McGlynn} {et~al.}(2012)}]{mcglynn_2012}
{McGlynn}, S. {et~al.} 2012, in `Gamma-Ray Burst 2012', Munich, May 7-11, 2012,
  eds. A. Rau and J. Greiner, PoS(GRB 2012)[012]

\bibitem[{{McKinney} \& {Uzdensky}(2012)}]{mckinney_2012}
{McKinney}, J.~C. \& {Uzdensky}, D.~A. 2012, \mnras, 419, 573

\bibitem[{{Meszaros} {et~al.}(1993){Meszaros}, {Laguna}, \&
  {Rees}}]{meszaros_1993}
{Meszaros}, P., {Laguna}, P., \& {Rees}, M.~J. 1993, \apj, 415, 181

\bibitem[{{M{\'e}sz{\'a}ros} {et~al.}(2002){M{\'e}sz{\'a}ros}, {Ramirez-Ruiz},
  {Rees}, \& {Zhang}}]{meszaros_2002}
{M{\'e}sz{\'a}ros}, P., {Ramirez-Ruiz}, E., {Rees}, M.~J., \& {Zhang}, B. 2002,
  \apj, 578, 812

\bibitem[{{M{\'e}sz{\'a}ros} \& {Rees}(2000)}]{meszaros_2000}
{M{\'e}sz{\'a}ros}, P. \& {Rees}, M.~J. 2000, \apj, 530, 292

\bibitem[{{Metzger} {et~al.}(1974){Metzger}, {Parker}, {Gilman}, {Peterson}, \&
  {Trombka}}]{metzger_1974}
{Metzger}, A.~E., {Parker}, R.~H., {Gilman}, D., {Peterson}, L.~E., \&
  {Trombka}, J.~I. 1974, \apjl, 194, L19

\bibitem[{{Mimica} \& {Aloy}(2010)}]{mimica_2010}
{Mimica}, P. \& {Aloy}, M.~A. 2010, \mnras, 401, 525

\bibitem[{{Nakar} {et~al.}(2009){Nakar}, {Ando}, \& {Sari}}]{nakar_2009}
{Nakar}, E., {Ando}, S., \& {Sari}, R. 2009, \apj, 703, 675

\bibitem[{{Nakar} \& {Piran}(2005)}]{nakar_2005}
{Nakar}, E. \& {Piran}, T. 2005, \mnras, 360, L73

\bibitem[{{Narayan} {et~al.}(2011){Narayan}, {Kumar}, \&
  {Tchekhovskoy}}]{narayan_2011}
{Narayan}, R., {Kumar}, P., \& {Tchekhovskoy}, A. 2011, \mnras, 416, 2193

\bibitem[{{Nava} {et~al.}(2011){Nava}, {Ghirlanda}, {Ghisellini}, \&
  {Celotti}}]{nava_2011}
{Nava}, L., {Ghirlanda}, G., {Ghisellini}, G., \& {Celotti}, A. 2011, \aap,
  530, A21

\bibitem[{{Nava} {et~al.}(2012){Nava}, {Salvaterra}, {Ghirlanda}, {Ghisellini},
  {Campana}, {Covino}, {Cusumano}, {D'Avanzo}, {D'Elia}, {Fugazza}, {Melandri},
  {Sbarufatti}, {Vergani}, \& {Tagliaferri}}]{nava_2012}
{Nava}, L., {Salvaterra}, R., {Ghirlanda}, G., {et~al.} 2012, \mnras, 421, 1256

\bibitem[{{Paczynski}(1986)}]{paczynski_1986}
{Paczynski}, B. 1986, \apjl, 308, L43

\bibitem[{{Pe'er}(2008)}]{peer_2008}
{Pe'er}, A. 2008, \apj, 682, 463

\bibitem[{{Pe'er} {et~al.}(2006){Pe'er}, {M{\'e}sz{\'a}ros}, \&
  {Rees}}]{peer_2006}
{Pe'er}, A., {M{\'e}sz{\'a}ros}, P., \& {Rees}, M.~J. 2006, \apj, 642, 995

\bibitem[{{Pe'er} {et~al.}(2007){Pe'er}, {Ryde}, {Wijers}, {M{\'e}sz{\'a}ros},
  \& {Rees}}]{peer_2007}
{Pe'er}, A., {Ryde}, F., {Wijers}, R.~A.~M.~J., {M{\'e}sz{\'a}ros}, P., \&
  {Rees}, M.~J. 2007, \apjl, 664, L1

\bibitem[{{Pe'er} {et~al.}(2012){Pe'er}, {Zhang}, {Ryde}, {McGlynn}, {Zhang},
  {Preece}, \& {Kouveliotou}}]{peer_2012}
{Pe'er}, A., {Zhang}, B.-B., {Ryde}, F., {et~al.} 2012, \mnras, 420, 468

\bibitem[{{Piran}(1999)}]{piran_1999}
{Piran}, T. 1999, \physrep, 314, 575

\bibitem[{{Popham} {et~al.}(1999){Popham}, {Woosley}, \& {Fryer}}]{popham_1999}
{Popham}, R., {Woosley}, S.~E., \& {Fryer}, C. 1999, \apj, 518, 356

\bibitem[{{Preece} {et~al.}(1998){Preece}, {Briggs}, {Mallozzi}, {Pendleton},
  {Paciesas}, \& {Band}}]{preece_1998}
{Preece}, R.~D., {Briggs}, M.~S., {Mallozzi}, R.~S., {et~al.} 1998, \apjl, 506,
  L23

\bibitem[{{Preece} {et~al.}(2000){Preece}, {Briggs}, {Mallozzi}, {Pendleton},
  {Paciesas}, \& {Band}}]{preece_2000}
{Preece}, R.~D., {Briggs}, M.~S., {Mallozzi}, R.~S., {et~al.} 2000, \apjs, 126,
  19

\bibitem[{{Rees} \& {Meszaros}(1994)}]{rees_1994}
{Rees}, M.~J. \& {Meszaros}, P. 1994, \apjl, 430, L93

\bibitem[{{Rees} \& {M{\'e}sz{\'a}ros}(2005)}]{rees_2005}
{Rees}, M.~J. \& {M{\'e}sz{\'a}ros}, P. 2005, \apj, 628, 847

\bibitem[{{Ryde}(2004)}]{ryde_2004}
{Ryde}, F. 2004, \apj, 614, 827

\bibitem[{{Ryde}(2005)}]{ryde_2005}
{Ryde}, F. 2005, \apjl, 625, L95

\bibitem[{{Ryde} {et~al.}(2010){Ryde}, {Axelsson}, {Zhang}, {McGlynn}, {Pe'er},
  {Lundman}, {Larsson}, {Battelino}, {Zhang}, {Bissaldi}, {Bregeon}, {Briggs},
  {Chiang}, {de Palma}, {Guiriec}, {Larsson}, {Longo}, {McBreen}, {Omodei},
  {Petrosian}, {Preece}, \& {van der Horst}}]{ryde_2010}
{Ryde}, F., {Axelsson}, M., {Zhang}, B.~B., {et~al.} 2010, \apjl, 709, L172

\bibitem[{{Ryde} {et~al.}(2012)}]{ryde_2012}
{Ryde}, F. {et~al.} 2012, in `Gamma-Ray Burst 2012', Munich, May 7-11, 2012,
  eds. A. Rau and J. Greiner, PoS(GRB 2012)[011]

\bibitem[{{Sari} {et~al.}(1998){Sari}, {Piran}, \& {Narayan}}]{sari_1998}
{Sari}, R., {Piran}, T., \& {Narayan}, R. 1998, \apjl, 497, L17

\bibitem[{{Shemi} \& {Piran}(1990)}]{shemi_1990}
{Shemi}, A. \& {Piran}, T. 1990, \apjl, 365, L55

\bibitem[{{Spruit} {et~al.}(2001){Spruit}, {Daigne}, \&
  {Drenkhahn}}]{spruit_2001}
{Spruit}, H.~C., {Daigne}, F., \& {Drenkhahn}, G. 2001, \aap, 369, 694

\bibitem[{{Tchekhovskoy} {et~al.}(2010){Tchekhovskoy}, {Narayan}, \&
  {McKinney}}]{tchekhovskoy_2010}
{Tchekhovskoy}, A., {Narayan}, R., \& {McKinney}, J.~C. 2010, \na, 15, 749

\bibitem[{{Thompson}(1994)}]{thompson_1994}
{Thompson}, C. 1994, \mnras, 270, 480

\bibitem[{{Vlahakis} \& {K{\"o}nigl}(2003)}]{vlahakis_2003}
{Vlahakis}, N. \& {K{\"o}nigl}, A. 2003, \apj, 596, 1080

\bibitem[{{Vurm} {et~al.}(2011){Vurm}, {Beloborodov}, \&
  {Poutanen}}]{vurm_2011}
{Vurm}, I., {Beloborodov}, A.~M., \& {Poutanen}, J. 2011, \apj, 738, 77

\bibitem[{{Zalamea} \& {Beloborodov}(2011)}]{zalamea_2011}
{Zalamea}, I. \& {Beloborodov}, A.~M. 2011, \mnras, 410, 2302

\bibitem[{{Zdziarski} \& {Lamb}(1986)}]{zdziarski_1986}
{Zdziarski}, A.~A. \& {Lamb}, D.~Q. 1986, \apjl, 309, L79

\bibitem[{{Zhang} \& {Pe'er}(2009)}]{zhang_2009}
{Zhang}, B. \& {Pe'er}, A. 2009, \apjl, 700, L65

\bibitem[{{Zhang} \& {Yan}(2011)}]{zhang_2011}
{Zhang}, B. \& {Yan}, H. 2011, \apj, 726, 90

\bibitem[{{Zhang} {et~al.}(2011){Zhang}, {Zhang}, {Liang}, {Fan}, {Wu},
  {Pe'er}, {Maxham}, {Gao}, \& {Dong}}]{zhang_bb_2011}
{Zhang}, B.-B., {Zhang}, B., {Liang}, E.-W., {et~al.} 2011, \apj, 730, 141

\end{thebibliography}

\end{document}